\documentclass[%
 reprint,
 amsmath,amssymb,
 aps, 
]{revtex4-1}

\usepackage{graphicx}
\usepackage{dcolumn}
\usepackage{bm}
\usepackage{xcolor}
\usepackage{xr} 
\usepackage{subcaption}
\usepackage[justification=RaggedRight]{caption}

\usepackage{hyperref}

\begin{document}

\title{Field-induced phase transitions in ferro-antiferromagnetic diblock copolymers}

\author{Alberto Raiola$^1$, Emanuele Locatelli$^1$, Davide Marenduzzo$^2$ and Enzo Orlandini$^1$ \\
$^1$  Dipartimento di Fisica e Astronomia and Sezione INFN, Universit\`a di Padova, \\ Via Marzolo 8, I-35131 Padova, Italy \\
$^2$ SUPA, School of Physics and Astronomy, University of Edinburgh, Peter Guthrie Tait Road, Edinburgh, EH9 3FD, UK \\}

\begin{abstract}
We study the equilibrium properties of a model of magnetic diblock copolymer where each monomer is decorated with an Ising-like spin. Spins interact ferromagnetically within each block and antiferromagnetically across blocks, generating frustration between magnetic ordering and spatial organization.  By employing a mean-field approach and Monte Carlo simulations for self-avoiding walks on the cubic lattice, we investigate the system's response to an external magnetic field. We discover a rich phase diagram that includes: a swollen phase with both filaments magnetically disordered and spatially extended; a mixed compact phase characterized by a single globule in which the two filaments are strongly intertwined; a segregated compact phase composed of two globular, magnetically ordered, and spatially separated blocks. Further, if the magnitude of the intra-block ferromagnetic interaction differs between the two blocks, we observe a hybrid segregated (``tadpole'') phase where one extended block coexists with a collapsed one.
Mean-field predictions are in quantitative agreement with Monte Carlo results for the location of the phase boundaries. These findings provide a minimal statistical-mechanical framework for field-controlled self-assembly of tunable patterns by magnetically heterogeneous polymers. They may also serve as a simple platform to investigate the coupling between internal epigenetic-like states and chromatin folding.
\end{abstract}

\maketitle

\section{Introduction}
Magnetic polymers are composite soft materials that have garnered attention due to their equilibrium behaviour, resulting from the interplay between configurational entropy and magnetic interactions, as each monomer carries a magnetic moment or spin~\cite{rajca2001magnetic,boudouris2024spin,blundell2007molecular}. This interplay bestows them with interesting properties, such as reactivity and memory, making them candidates for smart materials. In fact it is possible to affect the conformation of a magnetic polymer with an external magnetic field, inducing a mechanical transformation; from a theoretical perspective, the characteristic temperature-induced transition between compact and swollen conformations at zero magnetic field has a first-order character~\cite{garel1999phase}, opening up the possibility for information storage~\cite{cavallini2005magnetic}. This is in contrast with conventional polymers whose collapse transition, called $\theta$-collapse, has a second-order nature~\cite{lifshitz1978some}. 
In a broader sense, polymers with magnetic-type degrees of freedom are examples of interacting systems where the competition between the spatial organization of the chain and magnetic interactions can produce various conformational phases and phase transitions that have been studied recently using different models~\cite{papale2018ising,foster2021critical,rudra2023critical}.

Besides their theoretical and practical technological relevance \cite{thevenot2013magnetic, miller2014organic, kalia2014magnetic}, magnetic polymers have received attention because of their potential insights into chromatin organization. In particular, a magnetic polymer elegantly models the interplay between the kinetics of the spreading of epigenetic marks and the conformational changes due to the substrate's folding~\cite{coli2019magnetic, michieletto2016polymer,michieletto2019nonequilibrium,jost2018,owen2023}.

In a statistical mechanical approach to the problem, one can model a magnetic polymer as a self-avoiding walk whose monomers carry spin variables, e.g. Ising spins $S_{i} = \pm 1$. These spins interact with each other via a suitable Hamiltonian when they are nearest neighbours in 3D space. The first theoretical study~\cite{garel1999phase} considered the standard ferromagnetic Ising Hamiltonian, but more complex models, such as Potts~\cite{coli2019magnetic, nakanishi2024emergence} or Blume-Emery-Griffiths (BEG)~\cite{raiola2025equilibrium} can be considered too.

In this work we introduce an additional layer of complexity to these models by adding magnetic heterogeneity along the chain. More precisely, we consider a model of magnetic diblock copolymers, i.e., chains made up of two linear blocks (A and B) chemically connected end-to-end. Each monomer is decorated with an Ising-like spin variable, and we consider ferromagnetic interactions among monomers within the same block (A-A and B-B interactions) and antiferromagnetic interactions among monomers in different blocks (A-B interaction). Such competing couplings may arise in when the different blocks are made of chemically distinct magnetic species, where intra-species interactions favor alignment while inter-species interactions can be engineered -- e.g., via surface functionalisation -- to favor anti-alignment. At a coarse-grained level, this situation is analogous to a binary Lennard–Jones mixture with attractive like-like and unfavorable cross interactions, here translated into competing magnetic couplings. This choice introduces magnetic frustration that competes with conformational entropy and polymer connectivity.

Our results are obtained by a mean field (analytically tractable) approximation of the model and Monte Carlo simulations of the system implemented on the cubic lattice to capture potential deviations from the mean field predictions. Our analysis reveals that the frustration and competition between blocks due to the antiferromagnetic coupling transform the system into a responsive material, exhibiting a rich phase diagram and phase transitions. Specifically, we observe a transition from a mixed phase, characterized by highly intertwined polymer blocks, to a segregated phase where the two filaments become spatially distinct. Notably, this transition is triggered by the application of an external magnetic field. Furthermore, by modulating the strength of ferromagnetic interactions within each block,  more exotic phases can emerge, such as a 'tadpole' phase where a globular and an extended phase coexist in space.\\

The manuscript is organised as follows: in section \ref{sec:model} we introduce the model, discussing the choice of the magnetic interactions in more detail; in section~\ref{sec:MeanField} we determine the free energy and the phase diagrams of the model in the mean-field approximation using the Bragg-Williams method. Further, in section~\ref{sec:MonteCarlo} we discuss the details of the Monte Carlo simulations, report the results either at constant temperature (Sec.~\ref{sec:MCT}) or constant magnetic field (Sec.~\ref{sec:MCH}) and compare them with the mean-field predictions; the last section~\ref{sec:Discuss} is devoted to discussions and conclusions.

\section{Model and Mean Field theory}
\label{sec:model}
\begin{figure}[]
    \centering
    \includegraphics[width=1.\linewidth]{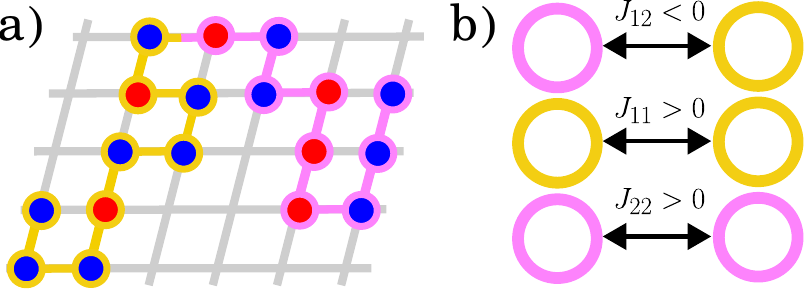}
    \caption{(a): Cartoon of the magnetic diblock copolymer model on the square lattice. The polymer is a self-avoiding walk (SAW) $\gamma$ and is composed of two blocks $\gamma_{1}$ and $\gamma_{2}$, highlighted in yellow and pink. Each vertex (monomer) is decorated with a spin variable $S_{i}$, with possible values $S_{i} = 1$ (blue) and $S_{i} = -1$ (red). (b) Summary of the interactions among spin variables: there is an antiferromagnetic coupling between pairs of spins belonging to different blocks and a  ferromagnetic one between pairs of spins in the same block.}
    \label{fig:Cartoon}
\end{figure}

We consider the ensemble $\{\gamma\}$ of self-avoiding walks (SAW) of $2N -1$ steps on a lattice of coordination number $z$. Each instance $\gamma$ of the ensemble represents a spatial configuration of the diblock copolymer, composed of $2N$ monomers, and is characterised by the adjacency matrix $\Delta_{i,j}^{\gamma}$. By definition $\Delta_{i,j}^{\gamma}=1$ if the pair of vertices of the walk,  $(i,j)$ are neighbors in space and $\Delta_{i,j}^{\gamma}=0$ otherwise. Clearly $\Delta_{i,i+1}^{\gamma}=1,\quad i=1,2N-1$.\\
Each block is a self-avoiding walk, $\gamma_{1}$ and $\gamma_{2}$, such that $\gamma = \gamma_{1} \cup \gamma_{2}$. We show a cartoon of this system in Fig.~\ref{fig:Cartoon}(a): the blocks of the polymer are colored in yellow and pink and the spin variables are colored in blue ($S_{i}=+1$) or red ($S_{i}=-1$). 

Neighbouring spins are coupled by an Ising-like interaction term; given a SAW $\gamma$, the Hamiltonian of the model can be written as:
\begin{align}
\mathcal H(\gamma, \{S\}) &= -\frac{J_{11}}{2} \sum_{ij \in \gamma_{1}}S_{i}\Delta_{ij}^{\gamma}S_{j} -\frac{J_{22}}{2} \sum_{ij \in \gamma_{2}}S_{i}\Delta_{i,j}^{\gamma}S_{j}  \cr
&+J_{12} \sum_{i \in \gamma_{1}} \sum_{j \in \gamma_{2}}S_{i}\Delta_{i,j}^{\gamma}S_{j} - H \sum_{i \in \gamma}S_{i},
\label{originalHamiltonian}
\end{align}
where $J_{11}$ and $J_{22}$ are the coupling constants of blocks 1 and 2 respectively, and $J_{12}$ is the cross-block counterpart. 
Notice that monomers belonging to the same block interact ferromagnetically (i.e., $J_{11}, J_{22} > 0$): as such, each of the two blocks behaves as an Ising magnetic polymer~\cite{garel1999phase}. 
Instead, monomers residing on different blocks interact antiferromagnetically with coupling constant $J_{12} > 0$. In Figure~\ref{fig:Cartoon}(b), we report a visual representation of these magnetic interactions. As  mentioned in the Introduction, the inter-block interaction, ruled by the term proportional to $J_{12}$ in~\eqref{originalHamiltonian},
brings in a competition between the inter- and intra-strand contacts. 
\subsection{Mean-Field free energy density}
\label{sec:MeanField}
In previous works on magnetic polymers, a common approach to derive the mean field free energy density of the model was based on the  Hubbard-Stratonovich transformation, followed by a homogeneous saddle point approximation~\cite{garel1999phase,raiola2025equilibrium}. Despite its elegance and rigour, this method requires the computation of the inverse of the adjacency matrix $\Delta_{i,j}^{\gamma}$ that, in the case of polymers with a diblock structure, is not a trivial task. Here, we employ the Bragg - William approximation (BW)~\cite{huang2009introduction}, a simpler and widely used method that works well in the presence of antiferromagnetic interactions ~\cite{Agra_2006}. 
\begin{figure*}[]
    \centering
      \includegraphics[width=0.9\linewidth]{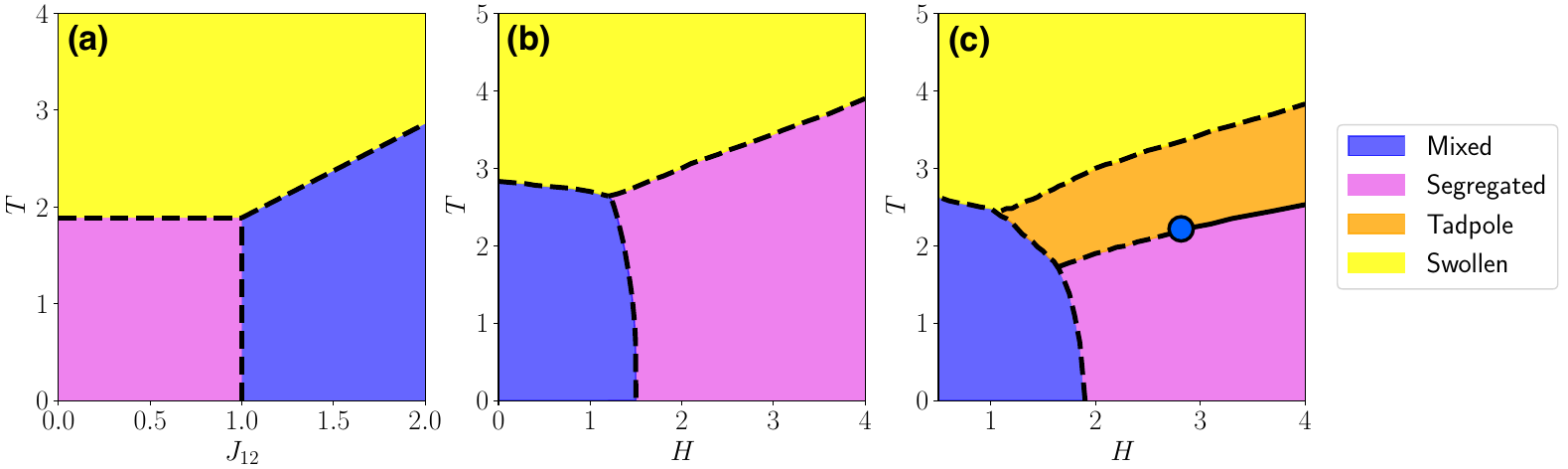}
     \includegraphics[width=0.9\linewidth]{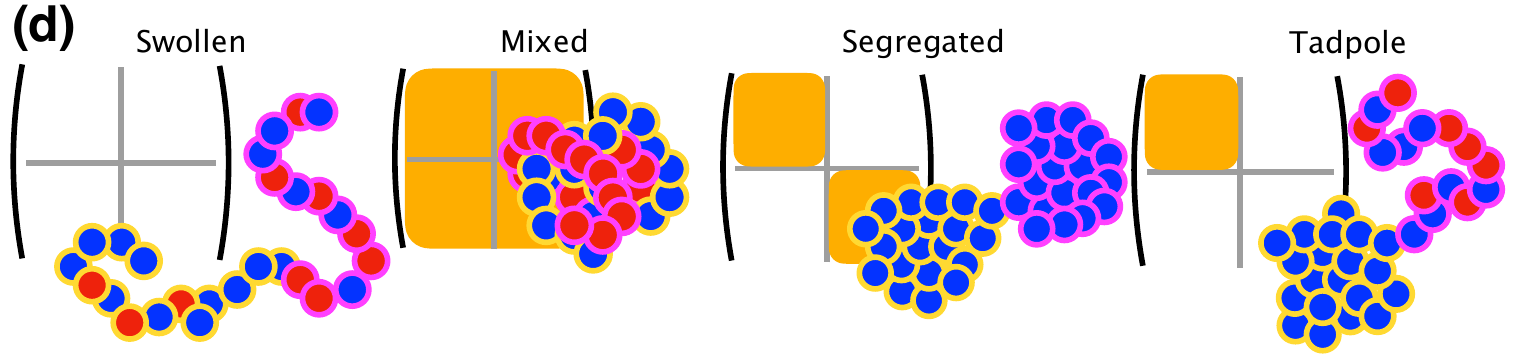}
    \caption{{Mean field phase diagrams of the magnetic diblock copolymer model}. In panels (a)-(c) three different section of the phase diagrams are shown: (a) $J_{12}-T$ plane at $H=0$ and $J_{11} = J_{22} = 1$, (b)-(c) $H-T$ plane for $J_{12} = 2$ and (b) $J_{11}=J_{22}=1$ or (c) $J_{11} = 1.0$ and $J_{22} = 0.5$. Four equilibrium phases are present: (i) the \emph{swollen phase} where both blocks are extended and magnetically disordered; (ii) the \emph{mixed phase}, characterized by a compact conformation where the two blocks are intertwined and have opposite magnetisation; (iii) the \emph{segregated phase}, characterized by two compact globules with no inter-filaments contacts and, at $H>0$, same-sign magnetisation; (iv) the \emph{asymmetrically segregated phase} (or \emph{tadpole phase}) \emph{phase}, that can be observed when $J_{11} \neq J_{22}$ and is characterized by the coexistence of one globular and one swollen block, depending on whichever block has the stronger ferromagnetic interaction. Dashed and solid lines denote, respectively, first-order and second-order phase transitions. The blue circle in panel (c) represents a tricritical point located at $H = 2.95$ and $T = 2.25 J$.
    Notably, if  $J_{12}< \max (J_{11}, J_{22})$ at $H=0$, the mixed phase does not occur. Panel (d) reports cartoons of typical conformations that can be observed in the different phases, as well as a qualitative graphical representation of their contact matrix. In the swollen phase (i), there is a vanishing number of contacts on the whole chain in the MF approximation. In the mixed phase (ii), under the Hamiltonian walk approximation, contacts are uniformly distributed on the chain. In the segregated phase (iii), contacts are mainly distributed within the two diagonal blocks -- the cartoon refers to the case $H > 0$, where $m_{1} m_{2} > 0$. Finally, in the tadpole phase (iv), contacts appear only within the collapsed block. 
    }
    \label{fig:Phases}
\end{figure*}
The detailed derivation of the MF free energy density is presented in Appendix.~\ref{sec:Fderivation}, while here we summarize the main findings and key arguments.\\
We aim to find the expression for the mean-field energy and entropy, that read
\begin{equation}
    E_{BW} = \langle \mathcal{H}(\gamma,\{S\})\rangle_p \qquad \text{and} \qquad \beta TS_{BW}=-\langle \log p\rangle_p
\end{equation}
respectively, where $\beta=(k_BT)^{-1}$ and the average is performed according to the probability $p(\{S\})$ of observing a certain spin configuration $S=\{S_{1}, S_{2}, \dots, S_{2N}\}$, in the BW approximation, and over all possible SAWs.
The mean-field entropy is found to be (see Sec.~\ref{sec:Fderivation})
\begin{align}
     \label{entropyMF}
     \beta TS_{BW} &= -\left \langle \log p \right \rangle_p \\
      &= -N\left ( \frac{1 + m_1}{2} \log \frac{1 + m_1}{2} + \frac{1 - m_1}{2}\log \frac{1 - m_1}{2}  
     \right ) \cr &- N\left (
     \frac{1 + m_2}{2} \log \frac{1 + m_2}{2} + \frac{1 - m_2}{2}\log \frac{1 - m_2}{2} 
     \right )\nonumber
\end{align} 
To get the mean-field energy, we could employ the Hamiltonian walk approximation~\cite{garel1999phase, raiola2025equilibrium} as an effective average of the contact matrix over the whole set of SAWs, namely $\sum_{i,j}\Delta_{i,j}^{\gamma}\approx N/(z\rho)$, where $\rho$ is the mean density of the walk.
Here, however, additional care should be taken due to the diblock nature of the walk.\\ 
In general, the density can attain different values in block $1$ and $2$ and we introduce $0 < \rho_{1} < 1$ and $ 0 < \rho_{2} < 1$ as the average density of block 1 and 2 respectively; one can expect $\rho_1 \neq \rho_2$ when the ferromagnetic interactions are asymmetric, that is, $J_{11} \neq J_{22}$. These two densities are related to the global density $\rho$ as
\begin{equation*}
\rho = \frac{\rho_{1} + \rho_{2}}{2}
\label{RelationsRho}
\end{equation*}
so that, when $\rho_{1} = \rho_{2} = 1$ both blocks are in the compact phase and $\rho=1$. Note, however, that the mere knowledge of $\rho_{1}$ and $\rho_{2}$  does not discriminate between a mixed phase and two (one for each block) segregated compact phases. This can be achieved by introducing a further parameter, $\rho_{mix}$, and splitting the contributions of the contact matrix into intra-block  ($C_{11}$ and $C_{22}$) and inter-block ($C_{12}$) terms, namely
\begin{align}
\label{eq:C12}
2C_{11} = \sum_{i,j \in \gamma_{1}} \Delta_{i,j}^{\gamma} &= N z \rho_{1} \left (1 - \frac{\rho_{mix}}{2} \right ) \cr
2C_{22} = \sum_{i,j \in \gamma_{2}} \Delta_{i,j}^{\gamma} &= N z \rho_{2} \left (1 - \frac{\rho_{mix}}{2} \right ) \\
C_{12} = \sum_{i \in \gamma_{1}, j \in \gamma_{2}} \Delta_{i,j}^{\gamma} &= N z \frac{\rho_{1} + \rho_{2}}{2} \frac{\rho_{mix}}{2}.  \nonumber
\end{align}
where, with $\gamma_1$ and $\gamma_2$ we denote the subwalk describing the block 1 and 2 respectively.
Notice that $2C_{11} + 2C_{22} + 2C_{12} =2 N z \rho$, as expected in the Hamiltonian walk ~\cite{garel1999phase}. 
When $\rho_{mix} = 0$, $C_{12} = 0$ and no inter-block contact occurs, signaling a segregated phase. Moreover, $2 C_{11} = N z \rho_{1}$ and $2C_{22} = N z \rho_{2}$, implying that the Hamiltonian walk approximation works for each block independently. 
On the other hand, if $\rho_{mix} = 1$ all three contributions in Eq.~\eqref{eq:C12} are finite and, if $\rho_{1} = \rho_{2} = \rho$, that is, when $J_{11} = J_{22}$, we find $2C_{11} = 2C_{22} = C_{12}$, meaning that contacts are uniformly distributed throughout the whole matrix $\Delta$, describing a mixed phase. 
Using this approximation, the mean-field energy reads:
\begin{widetext}
\begin{equation}
E_{BW} = -\frac{NJ_{11} m_1^2}{2} z \rho_{1} \left (1 - \frac{\rho_{mix}}{2} \right ) -\frac{N J_{22} m_2^2}{2}z \rho_{2} \left (1 - \frac{\rho_{mix}}{2} \right ) + NJ_{12}m_{1}m_{2}z \frac{\rho_{1} + \rho_{2}}{2}\frac{\rho_{mix}}{2} - NH(m_{1} + m_{2}).
\label{eq:Ebw}
\end{equation}
\end{widetext}
Next, we need to compute the entropic contribution of the chain $\beta F_{chain}$. This can be obtained from the enumeration of the self-avoiding walks of size $2N$ in a volume $V$.
When $\rho_{mix} = 1$ the diblock copolymer behaves as a single polymer. In this case~\cite{garel1999phase}:
\[
 \beta F_{chain}^{mix} = {2N} \log \frac{z}{e} - 2N\frac{1 - \rho}{\rho}\log (1 - \rho)
\]
On the other hand, when $\rho_{mix} = 0$ the two blocks are mutually independent, hence
\begin{align*}
\beta F_{chain}^{seg} &= {2N} \log \frac{z}{e} - N\frac{1 - \rho_{1}}{\rho_{1}}\log (1 - \rho_{1}) \\
&= - N\frac{1 - \rho_{2}}{\rho_{2}}\log (1 - \rho_{2}).
\end{align*}
To take into account both contributions, we write $\beta F_{chain}$ as an interpolation of the previous two expressions, with $\rho_{mix}$ as the interpolating parameter:
\begin{equation}
\beta F_{chain} = \beta F_{chain}^{seg}(1 - \rho_{mix}) +  \beta F_{chain}^{mix} \cdot \rho_{mix}    
\end{equation}
Finally, the mean-field free energy density reads
\begin{equation}
\beta f_{BW} = \frac{\beta F_{BW}}{2N} =  \frac{1}{2N} (\beta E_{BW} - \beta TS_{BW} - \beta F_{chain}).
\label{freeE}
\end{equation}
We highlight that the physics of the magnetic diblock copolymer is ruled by the self-consistent equations, reported in Sec.~\ref{sec:Fderivation}. In particular, from Eq.~\eqref{MFRhomix}, one can deduce that $\rho_{mix}$ is a binary variable, that is, $\rho_{mix} \in \{0,1\}$; as such, we expect to see the emergence of mixed and segregated phases. %
\subsection{Phase diagrams}
One can draw the mean-field phase diagram of the system by fixing the coupling constants and solving the self-consistent equations with respect to the order parameters. We showcase the phenomenology in Fig.~\ref{fig:Phases}; in general, we distinguish four phases 
\begin{itemize}
   \item A \emph{swollen} phase at high temperature, where both blocks are extended ($\rho_1, \rho_2 \simeq 0$) and segregated ($\rho_{mix} = 0$).
   \item A \emph{mixed compact} phase, where  $\rho_{1}, \rho_2 > 0$ and  $\rho_{mix} = 1$, appearing when $J_{12} > \max\{J_{11}, J_{22}\}$. In this phase, we expect the two blocks to display opposite magnetization, i.e., $m_1 \cdot m_{2} < 0$. We further expect contacts to be uniformly distributed along the whole chain, as the two blocks interpenetrate each other in a single compact globule.
    \item A \emph{segregated compact} phase, where $\rho_{1}, \rho_2 > 0$ and \hbox{$\rho_{mix} = 0$} appear if $J_{12} < \max\{J_{11}, J_{22}\}$ or at sufficiently large values of $H$. In the former case, the two blocks can be considered independent since there are no contacts between them, and as such, the values of the magnetisation may or may not have the same sign. Instead, in the latter case we expect \hbox{$m_1 \cdot m_2 > 0$}. In both cases, the polymer should arrange into two non-interpenetrated globules with no inter-filament contacts. 
    \item A \emph{hybrid segregated phase} (or \emph{tadpole phase}), where $\rho_{mix} = 0$, and one of the two filaments is swollen, i.e., $\rho_1 > 0, \rho_2 \simeq 0$ if $J_{11} > J_{22}$ or vice-versa otherwise. Indeed, for symmetry reasons, this phase appears only when $J_{11} \neq J_{22}$.
\end{itemize}
In Figure~\ref{fig:Phases} we report the MF equilibrium phase diagrams projected on: (i) the $J_{12}-T$ plane at $H=0$ (Fig.~\ref{fig:Phases}a) and the $H-T$ plane for (ii) $J_{11}=J_{22}$  (Fig.~\ref{fig:Phases}b) and (iii) $J_{11} \neq J_{22}$ (Fig.~\ref{fig:Phases}c).
If $J_{11} = J_{22} = 1$ (Fig.~\ref{fig:Phases}a) three phases occur:  at low temperature and for $J_{12} < 1$, the segregated phase is stable, as the antiferromagnetic interaction is not strong enough and there is no energetic gain in establishing inter-blocks contacts. The mixed phase is energetically favored for $J_{12} > 0$, and the swollen phase dominates at high temperature. In the second case (Fig.~\ref{fig:Phases}b), where $J_{12} = 2$ and $J_{11} = J_{22} = 1$, we also observe three phases. The mixed phase is stable at low values of $H$ and $T$; increasing the magnetic field, the system undergoes a first-order phase transition to a segregated phase, whereas at sufficiently high $T$, the swollen phase dominates. Note that, in the segregated phase, the two filaments can be considered as two independent magnetic polymers. As such, the segregated-swollen transition line is equivalent to the compact-segregated line of the magnetic polymer~\cite{garel1999phase}. This means that there is a multicritical point separating first-order and second-order transition lines. Such a point occurs only on the transition line between the tadpole and the segregated phase at the coordinates $(H = 2.95, T= 2.25)$ and is marked with a blue circle in Fig. ~\ref{fig:Phases}(c).
Note that the tricritical point on the tadpole/swollen transition line is located at $(H = 5.9, T= 4.5)$ i.e. well outside the range of $H$ reported in the panel. The same situation occurs in panel ~\ref{fig:Phases}(b).  As we do not discuss the nature of the transition lines in this work, we will defer more in-depth discussions to future work.\\
Note that, as $H\to \infty$, the model becomes equivalent to a $\theta$-point, since all spins are aligned with the magnetic field. As such, the phase boundary between swollen and compact phases becomes flat and has the value $T=J_{11}z=J_{22}z$ (see Appendix)
In the third case (Fig.~\ref{fig:Phases}c), where $J_{12} = 2$, $J_{11} = 1.0$ and $J_{22} = 0.5$, we observe the appearance of the tadpole phase which, as mentioned, can appear only if $J_{11} \neq J_{22}$. 
Notice that the two filaments can be considered as two independent magnetic polymers also in the tadpole phase. This observation can explain the shape of the tadpole region: the lower/upper branches mark the compact-swollen transition for blocks with lower/higher interaction strengths, respectively. This suggests that the extension of the tadpole region will depend on the ratio between the interaction strengths as well as on the relative fraction of the two magnetic species on the polymer. As before, we mention in passing that there are two multicritical points, one on each branch. Also in this case, for $H \to \infty$, the mean-field model becomes equivalent to a $\theta$-point for both filaments and the phase boundaries become flat with values $T_{1}=J_{11}z$ and $T_{2}=J_{22}z$.
Figure~\ref{fig:Phases}d) showcases cartoons of the expected polymer configurations in the different phases, with the corresponding expected contact maps: (i) in the mixed phase, contacts should appear anywhere along the chain, (ii)-(iii) in the segregated or tadpole phase, the majority of the contacts occur within each block or in a single block, respectively. Finally, (iv), in the swollen phase, many fewer contacts are expected.

\section{Monte Carlo simulations}
\label{sec:MonteCarlo}

\begin{figure*}[]
    \centering
    \includegraphics[width=0.8\linewidth]{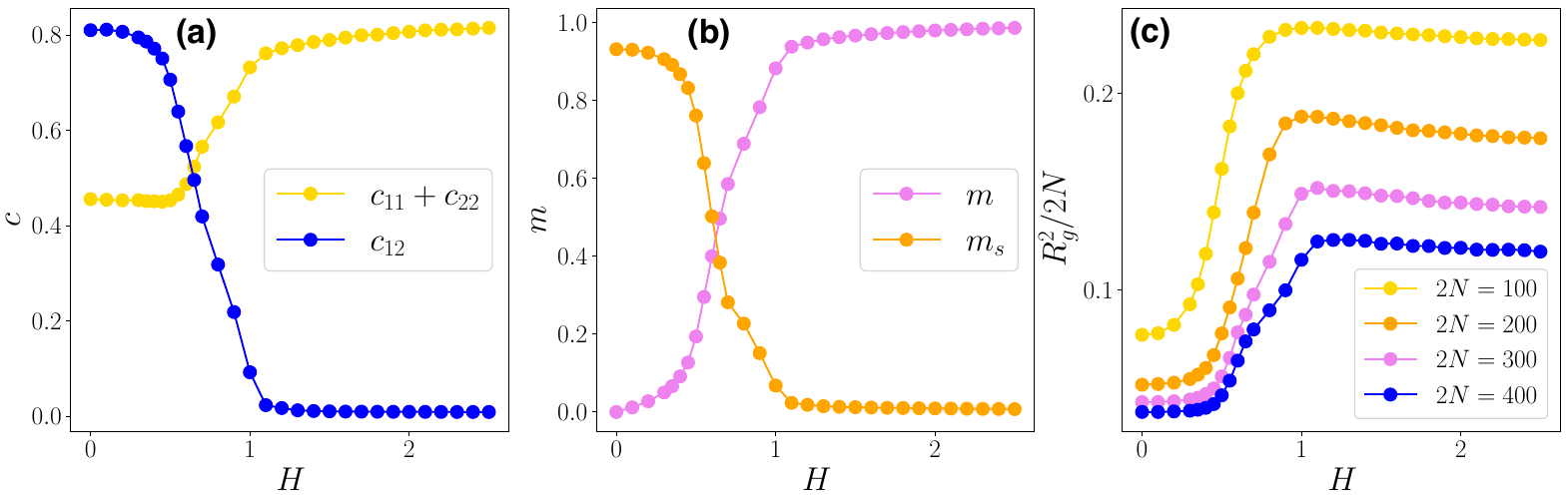}
     \includegraphics[width=0.9\linewidth]{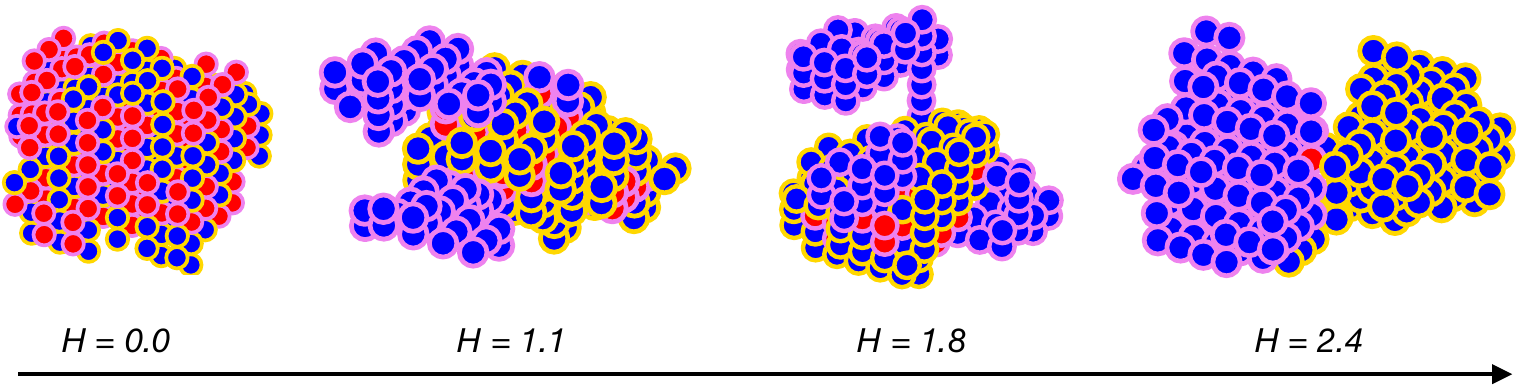}
    \caption{{Results of MC simulations at $T = 2.0$ and varying $H$ ($J_{11} = J_{22} = 1, J_{12} = 2$)}. (a) Inter-filament $c_{12}$ and intra-filament average number of contacts $c_{11} + c_{22}$ per monomer. (b) Magnetisation $m$ and staggered magnetisation $m_s$ per monomer. (c) Scaled mean square radius of gyration $R_g^2/2N$ for different values of $N$. Note that $c_{12}$ drops to zero for $H > 1.1$ while, concurrently, $c_{11} + c_{22}$ increases towards a plateau (a); a similar behavior holds for $m$ and $m_s$ (b),  indicating that the system is in a mixed antiferromagnetic compact phase for $H < 1.1$. No crossing appears in panel (c), suggesting that the system remains in a compact phase for all values of $H$ considered. The estimate of $H$ at the transition, $H^*$ is obtained by extrapolating the finite-size values $H_{max}(N)$, corresponding to the maximum of the variance of the inter-filament contacts, in the limit $N \to \infty$ . This gives $H^* \approx 1.0 \pm 0.1$.}
    \label{fig:MCT125}
\end{figure*}

To simulate the magnetic diblock copolymer in three dimensions, we consider the set of SAWs $\{ \gamma \}$ on a cubic lattice; each vertex carries a spin variable $S_{i} = \pm 1$. The set of configurations is sampled by a Monte Carlo algorithm, performing the following elementary moves: (i) a Glauber dynamics (or spin flip move) to update the spin configuration along the walk; (ii) a set of local Verdier-Stockmayer moves that increases the mobility of the Markov chain in proximity of the compact phase~\cite{Tesi}; (iii) pivot moves, crucial for the ergodicity of the system~\cite{madras1990monte}. Given a $2N$ diblock copolymer, one MC sweep comprises: (i) $1$ pivot move; (ii) $N/2$ local moves, and (iii) $2N$ spin flips move.\\

We consider Self-Avoiding Walks (SAWs) with lengths of  $100$, $200$, $300$, and $400$ vertices ($2N$). For each combination of parameters $J_{11}, J_{22}, J_{12}, H$ and $T$, we perform Monte Carlo (MC) simulations consisting of a minimum of $5\times 10^7$ sweeps.
Although a full exploration of the phase diagram of the system is beyond the scope of this study, we validate the mean-field prediction using two sets of simulations: one in which  $H$ varies at fixed $T$ (Sec.~\ref{sec:MCT}) and one in which $T$ varies at fixed $H$ (Sec.~\ref{sec:MCH}). These correspond to horizontal and vertical cuts in the $H-T$ plane, respectively.\\

To enhance sampling efficiency, we employ the multiple Markov chain algorithm (also known as replica exchange or parallel tempering) in both sets of simulations~\cite{Tesi}.

We run a number $n=25-30$ of Markov chains in parallel, either at different values of $H$ and a fixed temperature or at different values of $T$ at a fixed magnetic field. In the former case, the set of magnetic fields $H_{1}, H_{2}, \dots, H_{n}$ is such that $H_{i}$ and $H_{i + 1}$ are close enough that the energy distribution of the two chains has a significant overlap. A coupling between contiguous Markov chains is established, trying to swap configurations $i$ and $i+1$ with a probability equal to:
\[
p^{(H)}_{swap} = \min \left (1, e^{-\beta (m_{i+1} - m_i)(H_{i+1} - H_{i})} \right )
\]
In the latter case, the set of temperature values $T_{1}, T_{2}, \dots, T_{n}$ is chosen with the same rationale, and the swap probability is given by:
\[
p^{(T)}_{swap} =\min \left ( 1, e^{(\beta_{i+1} - \beta_i)(E_{i+1} - E_{i})} \right ) 
\]

We attempted a swap every $10^ {4}$ MC sweeps, such that the correlations between contiguous Markov chains are negligible; further details regarding the histogram of the swap probabilities and autocorrelation times of the observables can be found in the Appendix~\ref{AppendixDiagnostics}. 

We measure a set of observables describing the state of the diblock copolymer. The average number of contacts is defined as the average number of pairs of nearest-neighbour lattice nodes occupied by the polymer; by checking whether the neighbours belong to the same block or not, we distinguish between inter-filament contacts $C_{12}$ and intra-filament contacts $C_{11}$ and $C_{22}$. These quantities give insights on how much the system is mixed and whether the two blocks are collapsed or swollen; notice that, in MC simulations, one never expects contacts to vanish, even in the swollen phase. Furthermore, contacts along the backbone are counted, contributing $N - 1$ contacts in $C_{11}$ and $C_{22}$ and $2$ contacts in $C_{12}$. In what follows, we subtract these numbers by considering the average number of contacts per monomer as, $c_{11} = (C_{11}-(N-1))/(2N)$, $c_{22} = (C_{22}-(N-1))/(2N)$ and $c_{12} = (C_{12}-2)/(2N)$.  Finally, we call $c=c_{11} + c_{22}$ the number of intra-block contacts per monomer.

We monitor the magnetic properties of the system by computing the average magnetization per monomer within the two blocks, $m_{i} = \frac{1}{N} \left | \sum_{j \in \gamma_{i} }S_{j} \right |$ where $i = 1, 2$. The average magnetization per monomer of the whole system is $m = (m_{1} + m_{2})/2$; we also consider the staggered magnetization,  $m_{s} = |m_{1} - m_{2}|/2$. Since we consider only the symmetric case, that is, the two blocks have the same length, then $m_s>0$ signals the presence of a mixed phase in which the antiferromagnetic interaction is prevalent. 

Finally, we characterize the three-dimensional conformational properties of the polymer via the squared radius of gyration
\[
R_{g}^{2} = \frac{1}{2N} \left \langle \sum_{i = 1}^{2N} \|\vec R_{i} - \vec R_{cm} \|^{2} \right \rangle\,,
\]
where 
$\vec R_{cm} = \frac{1}{2N} \sum_{i = 1}^{2N} \vec R_{i}$ is the  centre of mass of the chain.
The $R_g^2$ of each block is similarly defined.
It is known that, as  $N \to \infty$,   $R_{g}^2  \sim N^{2\nu}$ where,  in three dimensions, $\nu = 1/3$ for compact conformations, $\nu = 1/2$ at the $\theta$ point and $\nu \simeq 3/5$ for a swollen polymer~\cite{vanderzande1998,guida1998critical,clisby_PRL_2010}. As we expect the polymer to transition between different scaling regimes crossing different phases, we highlight the finite size crossovers through the crossings of the curves $R_{g}^{2}/N^{2 \nu}$, with $\nu=1/2$. 
In addition we can employ the values of the temperature corresponding to such crossings, $T_c(N)$, to obtain an estimate of the transition temperature by plotting $T_c(N)$ against $1/N$, and extrapolate the value of the transition temperature $T^*$ in the $N\to\infty$ limit.\\
In the following, we present the results of the MC simulations performed at a fixed temperature (Sec.~\ref{sec:MCT}) and at a fixed magnetic field (Sec.~\ref{sec:MCH}). As mentioned, we selected specific vertical and horizontal cuts of the phase diagram; this will allow us to explore all the possible transitions driven by the magnetic field predicted by the mean-field approximation.

\begin{figure*}[]
    \centering
    \includegraphics[width=0.8\linewidth]{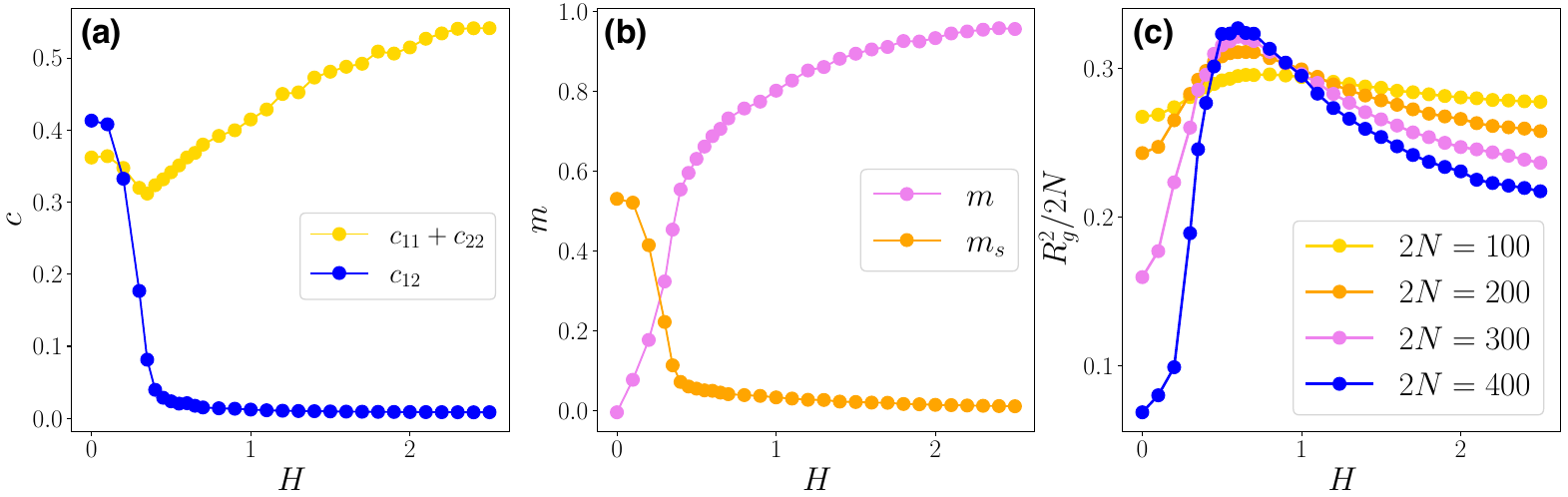}
     \includegraphics[width=0.9\linewidth]{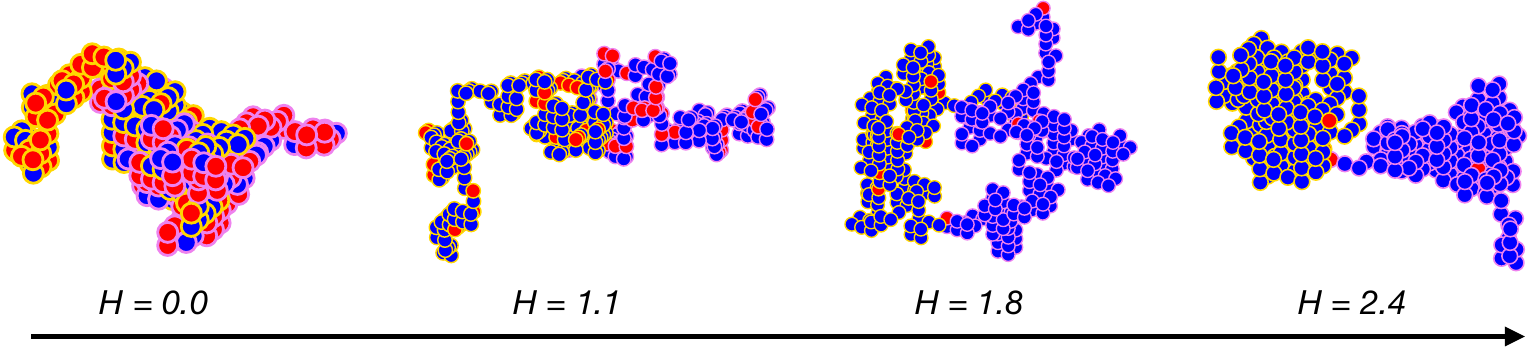}
    \caption{{Results of MC simulations at fixed $T = 2.5$ and varying $H$ ($J_{11} = J_{22} = 1, J_{12}=2$)}. (a) Inter-filament $c_{12}$ and intra-filament average number of contacts $c_{11} + c_{22}$ per monomer. (b) Magnetisation $m$ and staggered magnetisation $m_s$ per monomer. (c) Scaled mean square radius of gyration $R_g^2/2N$ for different values of $N$. At low values of $H$, $c_{12} > 0$ and $m_s>0$, confirming the presence of a mixed compact phase (see snapshot at $H=0.0$), while the presence of a minimum in the intra-filament contacts (a) suggests the onset of a more extended phase (see snapshots at $H=1.1,1.8$). In panel (c), the appearance of crossings between the curves of $R_g^2/N$ at different values of $N$ is revealing of two consecutive transitions: the first one between a compact mixed and a swollen disordered phase, and the second one between a swollen disordered and a compact segregated phase (see snapshot at $H=2.4$). The estimates of the magnetic field values at these transitions are $H_1^* = 0.52 \pm 0.03 $  and $H_2^* = 0.79 \pm 0.08 $ respectively.}
    \label{fig:MCT25}
\end{figure*}

\begin{figure*}[]
    \centering
    \includegraphics[width=0.8\linewidth]{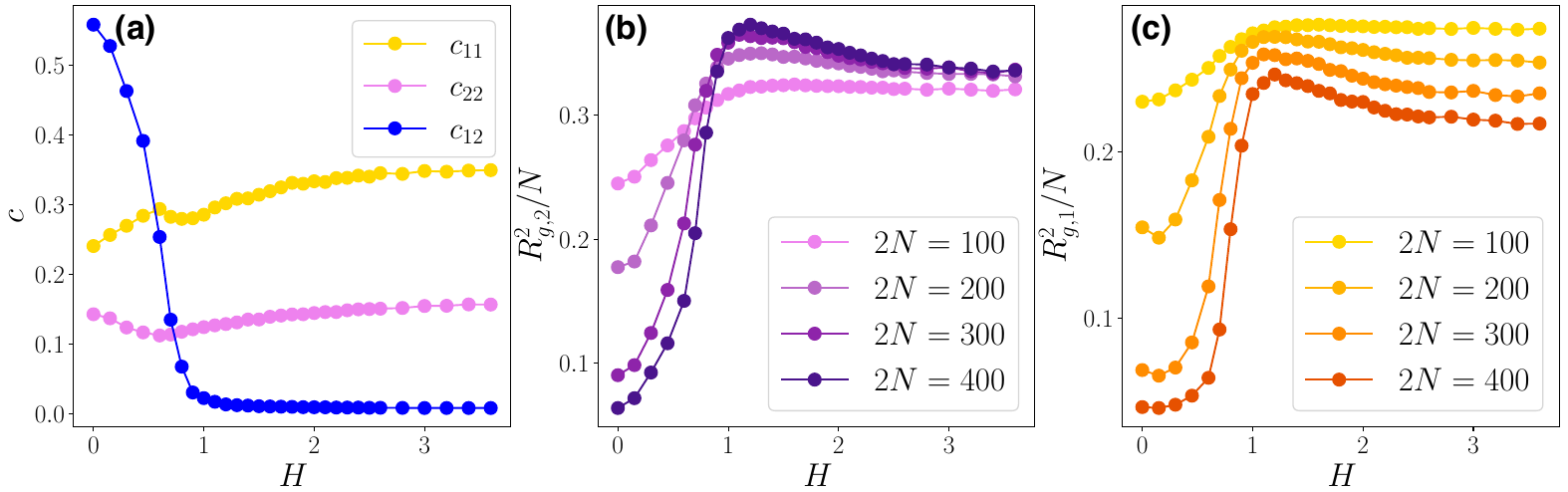}
    \includegraphics[width=0.8\linewidth]{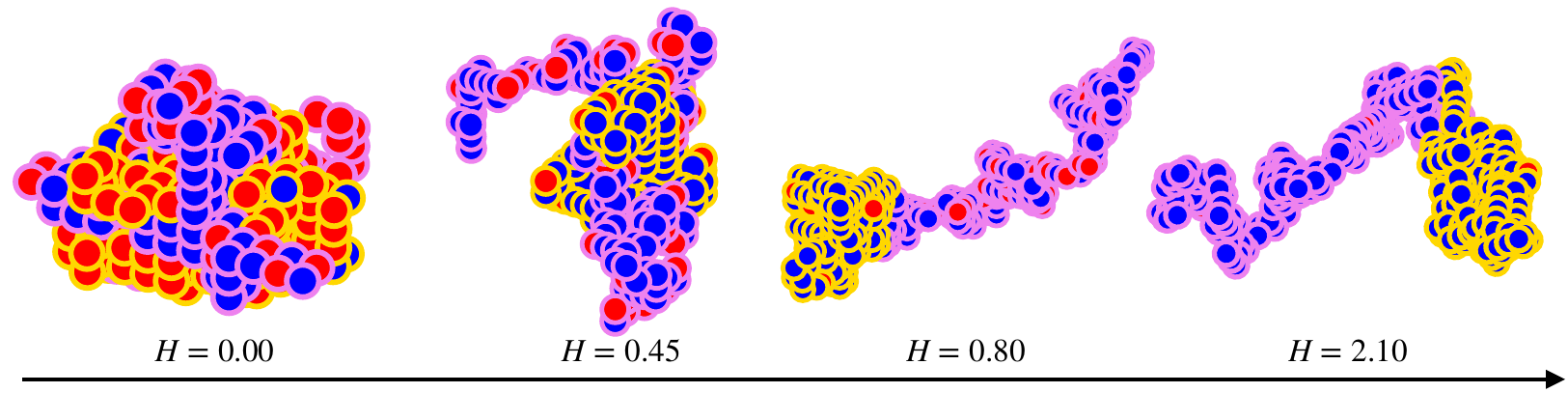}
    \caption{{Results of MC simulations at fixed $T = 2.5$ and varying $H$, for the asymmetric case $J_{11} \neq J_{22}$} ($J_{11} = 1, J_{22} = 0.5$ and $J_{12} = 2.0$). (a) Intra-filament and inter-filament average number of contacts per monomer $c_{11}$ and $c_{22}$, (b,c) Scaled mean square gyration radius of the second block $R_{g,2}^2/(N)$ (b)and of the first block $R_{g,1}^2/(N)$ (c) for different values of $N$. At low values of $H$, the observed large values of $c_{12}$ highlight a mixed compact phase (see snapshot at $H=0.00$). Notice that, since $J_{22} < J_{11}$, we have  $c_{22} < c_{11}$ for every value of $H$ considered. The crossings observed at $0.5 < H < 0.9$ of the scaled gyration radius curves of the second block (b)  pinpoint its conformational transitions from a compact to an extended phase  (see snapshots at $H = 0.45, 0.80$). In contrast, panel (c) suggests that the first block remains in a compact state for all values of $H$ considered (see snapshots). We extrapolate the value of the critical field at the transition to $H^* = 1.19 \pm 0.04$. }
    \label{fig:TADT225}
\end{figure*}

\subsection{Monte Carlo results at fixed temperature and variable magnetic field}
\label{sec:MCT}
We report here the results of the MC simulations in which we vary the magnetic field $H$ by keeping the temperature fixed at a given value. This corresponds to performing horizontal cuts in the $H-T$ plane (see the MF phase diagrams of Figures~\ref{fig:Phases}(b,c) as a reference). We focus on three cases: in (i) and (ii) we fix $J_{11} = J_{22} = 1, J_{12} = 2$ and (i) $T=2$ or (ii) $T=2.5$; in the last case (iii) we fix $J_{11} = 1$, $J_{22} = 0.5$, $J_{12} = 2$ and $T=2.5$. In all cases, we vary $H$ within the interval $0\leq H \leq 2.5$. With these choices, we expect to observe (i) the mixed-segregated transition, (ii) a double mixed-swollen-compact transition, and (iii) the tadpole phase.

\subsubsection{$J_{11} = J_{22} = 1, J_{12} = 2,  T = 2.0$}
\label{FixedT125}

The MC results, shown in Fig.~\ref{fig:MCT125}, reveal that there is a large number of inter-block contacts at small values of $H$; further, the two filaments display opposite magnetization, as expected from a mixed compact phase. A typical sampled configuration is visible in the leftmost snapshot, which refers to $H = 0$; colors are the same as in Fig. \ref{fig:Phases}. At $H=0$, the estimate of the Flory exponent is $\nu_{mix} \approx 0.34$, corroborating the validity of the Hamiltonian walk approximation used in the mean-field approach. As $H$ increases,  the inter-filament contacts and the staggered magnetization decrease to zero at $H\approx1.1$, while the intra-filament contacts and the total magnetization of the system increase (see Fig.~\ref{fig:MCT125}a,b). The other three snapshots showcase typical configurations of the system for $H\geq1.1$. Notice that the configuration sampled at the highest values of $H$ ($H=2.4$) is composed of two separate compact globules. For both globules, the estimation of the Flory exponent is $\nu_{sep} \approx 0.31$. 
The absence of crossings between the $R_{g}^{2}/N$ curves in Fig.~\ref{fig:MCT125} confirms that the system retains a compact conformation for the whole range of values of $H$ considered: the system transitions from a conformation characterised by a single, mixed globule and two separated globules. This confirms the MF prediction of the occurrence of a mixed-segregated transition for sufficiently high values of the magnetic field. Note that, in this case, the estimate of $H$ at the transition is obtained by extrapolating the location of the maxima of $Var(c_{12})$,$H_{max}(N)$, as $N\to\infty$.

\subsubsection{$J_{11} = J_{22} = 1, J_{12}=2, T = 2.5$}
\label{T250MC}
The MC results of Fig.~\ref{fig:MCT25} confirm the presence of a mixed phase. When $H \simeq 0$, the number of inter-filament contacts and the staggered magnetization per monomer are both greater than zero (see Fig.~\ref{fig:MCT25}a,b); a direct inspection of the typical conformations corroborates this finding (see the snapshot at $H=0.0$). Interestingly, the intra-filament contacts per monomer $c_{11}+c_{22}$ display a minimum for $H \simeq 0.4$, signaling the possible presence of an intermediate extended phase; this is also apparent from the snapshots at $H=1.1$ and $H=1.8$.
The presence of two sets of crossings between the $R_{g}^{2}/N$ curves allows us to pinpoint better the onset of a mixed-swollen transition as well as of a second one, which drives the system into a compact segregated phase (see also the snapshot at $H=2.4$). The first set of crossings, i.e., those related to the mixed-swollen transition, occurs in the range $0.3 < H < 0.5$ while the second set, corresponding to the swollen-segregated transition, is found within $0.8 < H < 1.1$. The $N\to\infty$ estimates of the critical values of the magnetic field $H^*$ for both transitions  are $H_1^* = 0.52 \pm 0.03$ and $H_2^* = 0.79 \pm 0.08$, respectively.
These findings further corroborate the MF predictions (see Fig.~\ref{fig:Phases}), regarding the presence of a mixed-swollen transition followed by a swollen-segregated one.
As could be expected, the temperature value at which this double transition is numerically observed ($T=2.5$) does not coincide with those suggested by the Mean Field (MF) phase diagram Fig.~\ref{fig:Phases}b  but is slightly lower.
Finally, note that both snapshots at $H=0.0$ and $H=2.4$ lie near the phase boundary and, therefore, reveal conformations that are less compact than their counterparts in Fig.~\ref{fig:MCT125}. 

\subsubsection{$J_{11} = 1, J_{22} = 0.5, J_{12} = 2.0, T = 2.5$}

The MC results, shown in Fig.~\ref{fig:TADT225}, highlight the presence of a mixed phase at low values of $H$, as the number of inter-filament contacts $c_{12}$ is greater than zero (see also the snapshot at $H=0.00$). Note that the number of intra-block contacts is different in the two blocks, as expected from imposing different values to $J_{11}$ and $J_{22}$; in particular, $c_{22}$ displays a minimum when $H = H^* \approx 0.8$, suggesting the onset of an extended phase (see Fig.~\ref{fig:TADT225}a). To better understand this transition, we examine the squared radius of gyration of the individual blocks $R^2_{g,1}$ and $R^2_{g,2}$ scaled by the block length $N$, see Figure~\ref{fig:TADT225}b,c). The crossings of the $R_{g,2}^2/(N)$ curves in Fig.~\ref{fig:TADT225}b) suggest a transition from a compact phase to an extended phase of the second filament, occurring in a range $0.5 < H < 0.9$.  By extrapolating the values of $H$ at which these crossings occur, we obtain the estimate  $H^* = 1.19 \pm 0.04$.
As $H$ increases, the curves at larger values of $N$ tend to collapse to a constant value, and no further crossings are observed. According to the mean-field picture, we should expect a swollen phase, implying that the values of $R_{g,2}/N$ curves should increase with increasing $N$. However, mean-field also predicts that, as $H$ increases, the system should become equivalent to a $\theta$-point, which implies a transition temperature $T_{\theta}\approx3.6$ with $J=1$. As $J_{22}=0.5$, the transition temperature for the second block becomes $T\approx1.9$, which is reasonably close to the chosen temperature $T=2.5$. As such, we observe a phenomenology consistent with an ideal polymer, due to finite size.
On the contrary, the rescaled curves of $R_{g,1}^2/(N)$ in Fig.~\ref{fig:TADT225}c) do not display any crossings, showing that the first block remains in the compact regime for the entire range of $H$ considered. This compact-to-extended phase transition of the second block is qualitatively confirmed by the snapshots of three typical configurations sampled at $H=0.45, 0.80, 2.10$.

In summary, these findings align with the MF prediction: when the two ferromagnetic coupling constants differ, a tadpole phase may arise; this is characterized by a compact to swollen transition for the filament with the smallest ferromagnetic interaction. Moreover, this transition takes place at intermediate values of the magnetic field and at sufficiently low temperatures.
\subsection{Monte Carlo results at fixed magnetic field and variable temperature}
\label{sec:MCH}
\begin{figure*}[]
    \centering
    \includegraphics[width=0.8\linewidth]{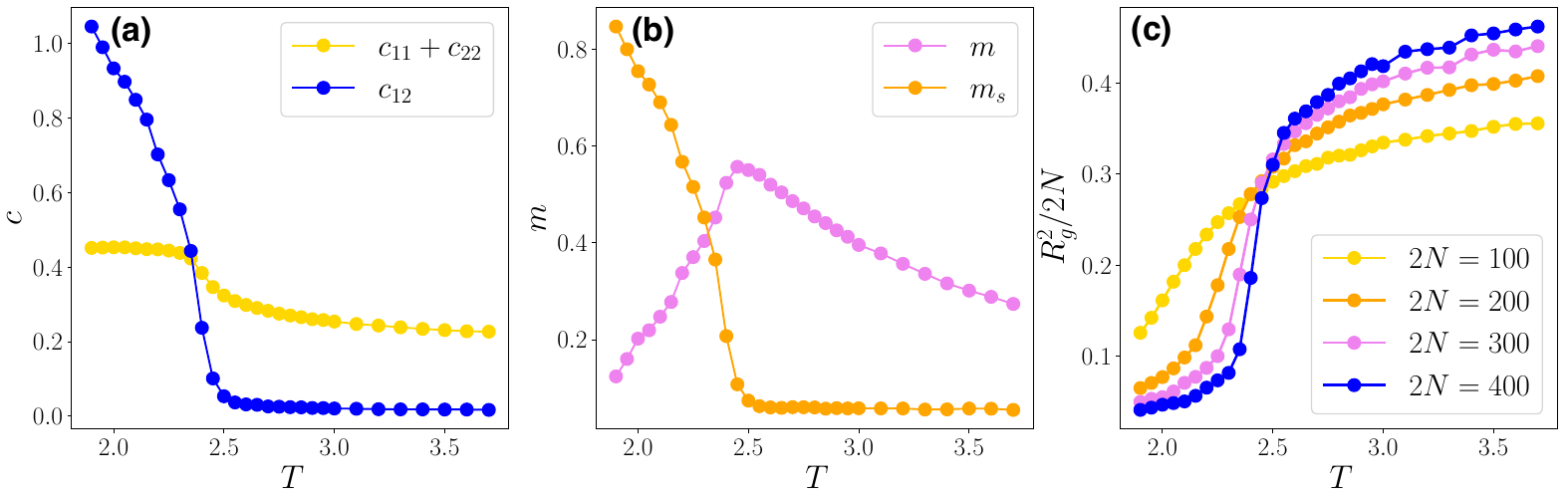}
     \includegraphics[width=0.8\linewidth]{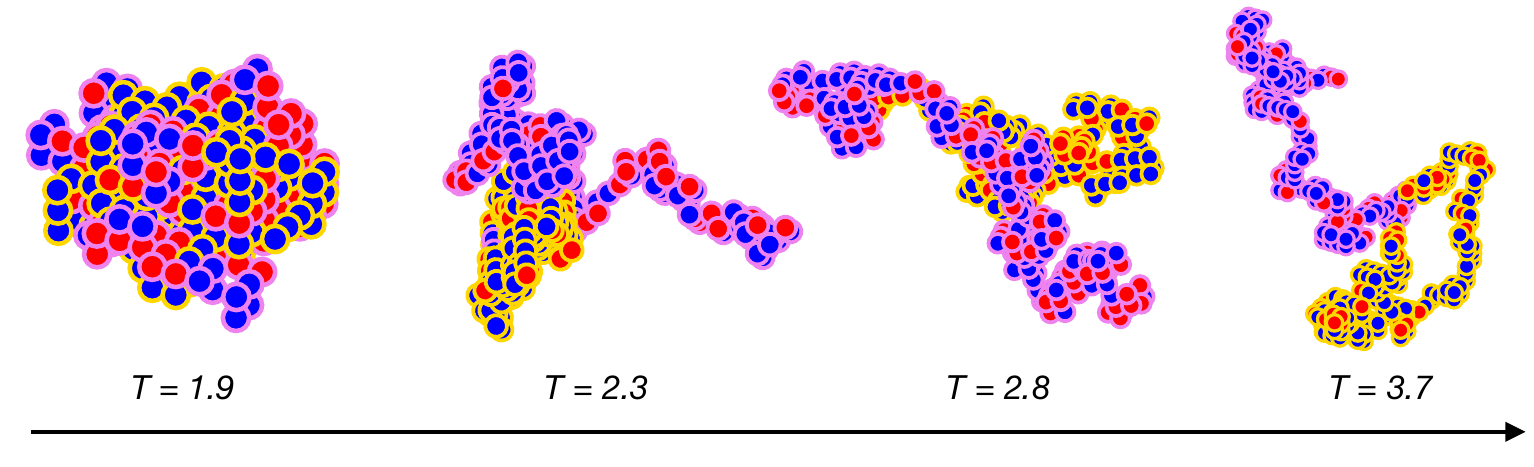}
    \caption{{Results of MC simulations at $H = 0.5$ and varying $T$ ($J_{11} = J_{22} = 1.0$, $J_{12} = 2$).} (a) Inter-filament and intra-filament average number of contacts per monomer, $c_{12}$ and  $c_{11} + c_{22}$, (b) magnetization $m$ staggered magnetization $m_s$ per monomer and (c) $R_g^2/(2N)$ as a function of $T$ for different values of $N$. 
    A transition from a mixed to a non-mixed phase is signaled by the vanishing of the staggered magnetisation at $T^* \approx 2.5$, accompanied by a maximum of the ordinary magnetization. The magnetisation displays a non-monotonic behaviour, since, for low values of $T$, the two filaments are anti-magnetized, while for large values of $T$ thermal fluctuations dominate, leading to the swollen, magnetically disordered, phase. As before, the crossings between the curves $R_g^2/N$ estimated at different values of $N$ highlight a phase transition from a mixed compact to a swollen phase; this transition is also qualitatively suggested by the snapshots of some typical configurations. We extrapolate the transition temperature from the crossing points as $T^* = 2.56 \pm 0.02$.}
    \label{fig:H05}
\end{figure*}
\begin{figure*}[]
    \centering
    \includegraphics[width=0.8\linewidth]{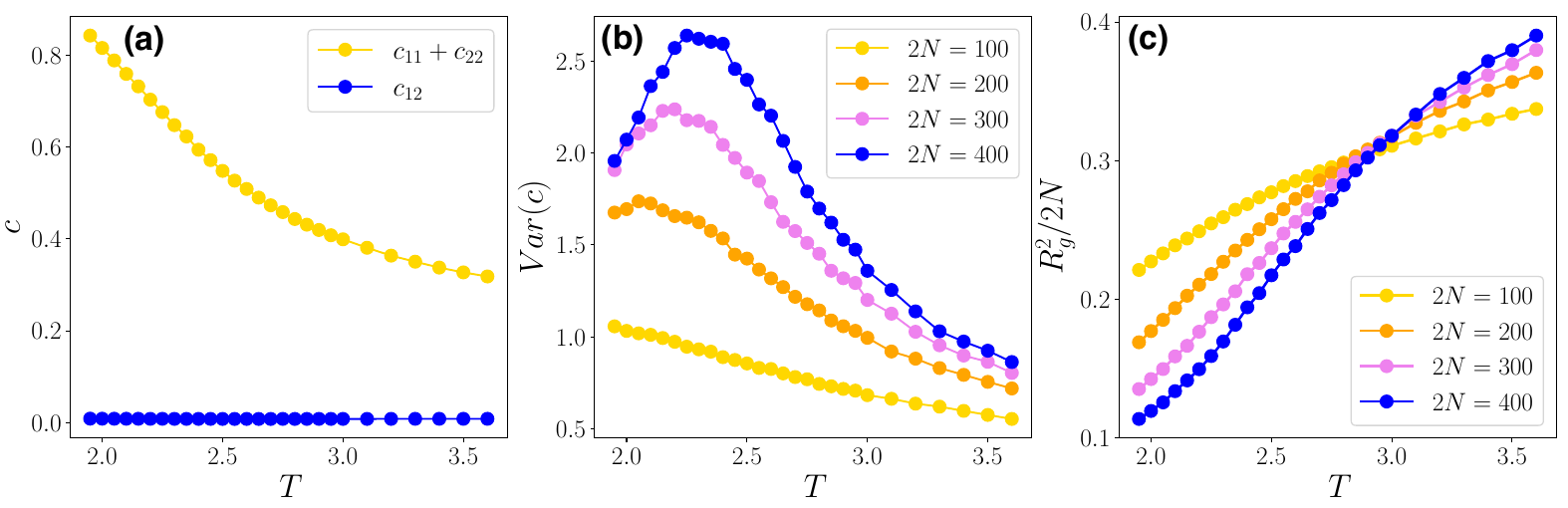}
     \includegraphics[width=0.8\linewidth]{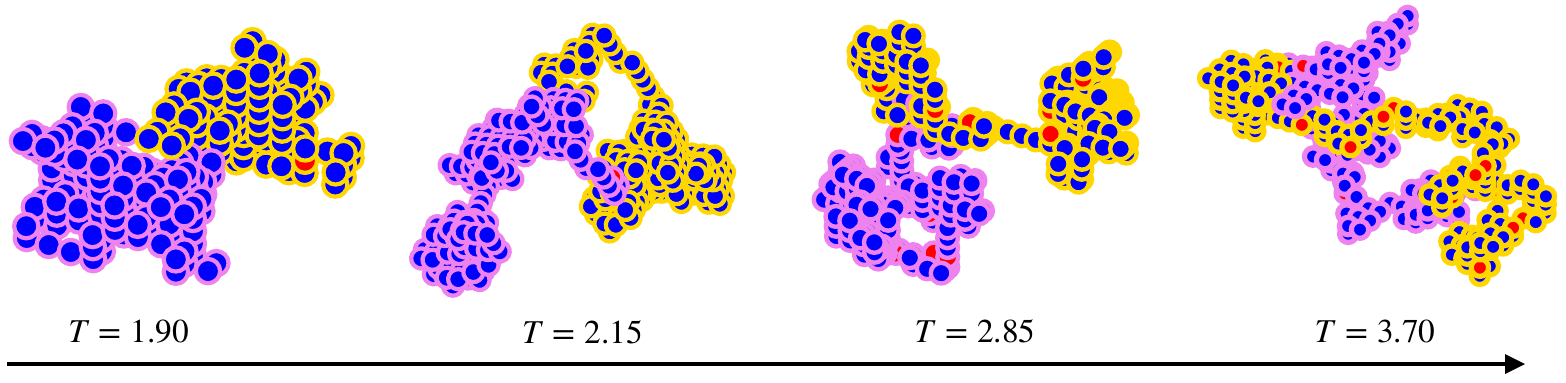}
    \caption{Results of the MC simulations at $H = 2.5$ and varying $T$ ($J_{11} = J_{22} =1$, $J_{12} = 2.0$). (a) Inter-filament and intra-filament average number of contacts per monomer $c_{12}$ and $c_{11} + c_{22}$, (b) variance of the number of contacts $Var(c)=Var(c_{11}+c_{22})$ and (c) scaled square radius of gyration, $R_g^2/(2N)$ for different values of $N$. Notice that $c_{12} \simeq 0$ for all values of $T$, namely, no mixed phase is observed. Moreover, the number of intra-filament contacts decreases with increasing $T$, suggesting conformational transition to an extended phase.   
    The presence of a compact (segregated) to swollen transition is indeed highlighted by the maximum of $Var(c)$ and by the intersections between the $R_g^2/N$ curves at different values of $N$; this transition is also suggested by the snapshots of some typical configurations sampled at different values of $T$. We estimate the transition temperature is  $T^* = 3.15 \pm 0.02$. 
    }
    \label{fig:H25}
\end{figure*}
\begin{figure*}[]
    \centering
    \includegraphics[width=0.8\linewidth]{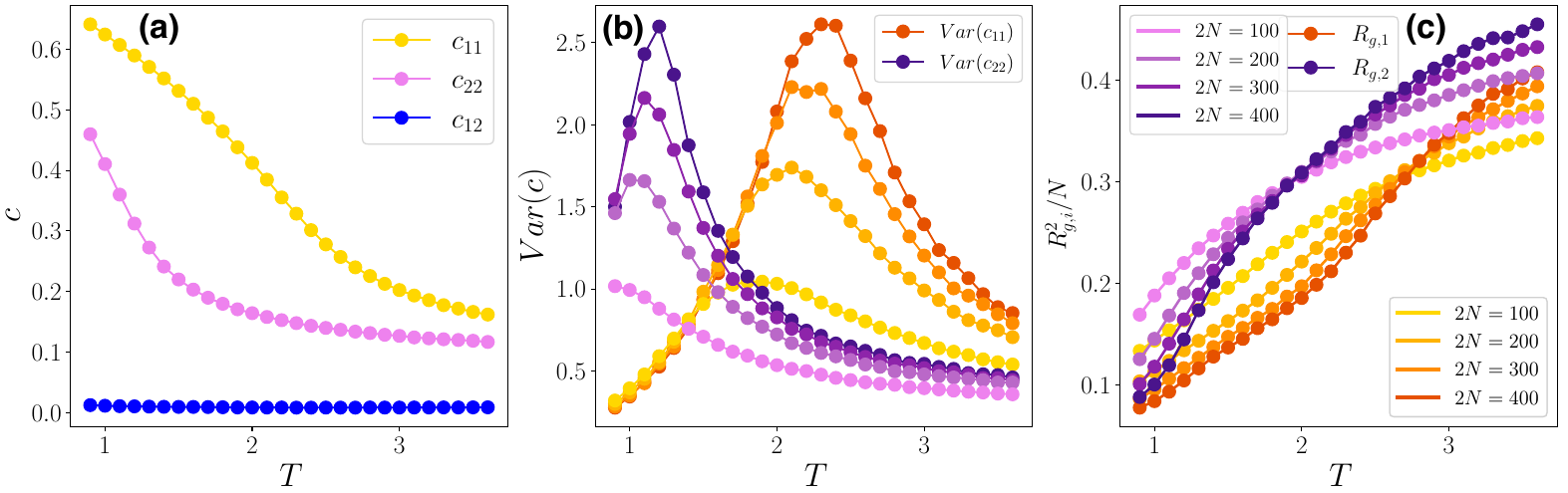}
     \includegraphics[width=0.8\linewidth]{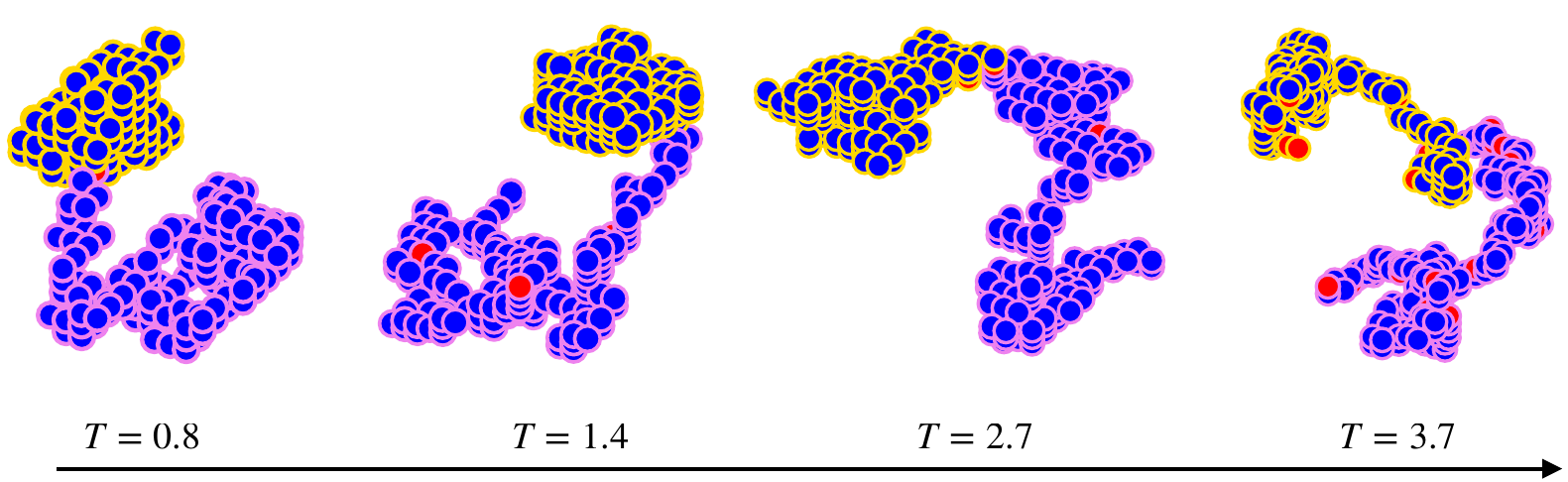}
    \caption{{Results of the MC simulations at fixed $H = 2.5$ and varying $T$, for the asymmetric case ($J_{11}\ne J_{22}$ ($J_{11} = 1$, $J_{22} = 0.5$ and  $J_{12} = 2$)}. (a) Inter-filament and intra-filament average number of contacts per monomer, $c_{12}$, $c_{11}$ and $c_{22}$, (b) variance of the number of contacts $Var(c)$ and (c) scaled mean square radius of gyration $R_{g,1}^2/N$ $R_{g,2}^2/N$ for different values of $N$. As in Fig.~\ref{fig:H25}, $c_{12} \simeq 0$; here, however, since $J_{22} = J_{11}/2< J_{11}$, $c_{22}$ decreases faster than $c_{11}$. The variance of both $c_{11}$ and $c_{22}$ presents a maximum increasing and moving towards higher temperatures as $2N$ increases. As expected the maxima of the second filament occur at lower temperatures than those of the first filament. By extrapolating in the $N\to\infty$ limit the location of the crossings of the  $R_{g,2}^2/N$ curves and those of the $R_{g,1}^2/N$ curves, we estimate the two transition temperatures as $T^*_1 = 3.0 \pm 0.2$ and $T^*_2 = 2.1 \pm 0.3$. 
    }
    \label{fig:TADH25}
\end{figure*}

MC simulations performed at a fixed value of the magnetic field and variable $T$ correspond to performing vertical cuts in the $H-T$ plane of the phase diagrams in Figure~\ref{fig:Phases}. As before we consider three cases: in (i) and (ii) we fix $J_{11} = J_{22} = 1, J_{12} = 2$ with $H=0.5$ in (i) and $H=2.5$ in  (ii); in case (iii) we fix $J_{11} = 1$, $J_{22} = 0.5$, $J_{12} = 2$ and $H=2.5$. In all cases, we vary $T$ within the interval $1.9\leq T \leq 3.7$. With these choices, we aim to  observe the compact-swollen transition and the tadpole phase

\subsubsection{$J_{11} = J_{22} =1 $, $J_{12} = 2$, $H = 0.5$}
 
The MC results, reported in Fig.~\ref{fig:H05}, show the presence of a compact-swollen phase transition, highlighted either by the trends of the number of intra-filament contacts (Fig.~\ref{fig:H05}a) and the magnetisation (Fig.~\ref{fig:H05}b) as a function of $T$. In particular, the behaviour of the staggered magnetisation suggests that the diblock copolymer assumes a compact-mixed conformation at low values of $T$; this is qualitatively confirmed by the snapshot of typical configurations sampled at $T=1.9$ and is in agreement with the results reported in Fig.~\ref{fig:MCT125}.
Further, we highlight that the average number of intra-filament contacts per monomer $c_{11} + c_{22}$ decreases with $T$, reaching a plateau at sufficiently high temperature. It is interesting to notice that the magnetization displays a nonmonotonic behaviour, with a maximum at $T^* \approx 2.5$. Indeed, the two blocks tend to assume opposite magnetization at low values of $T$ while the diblock copolymer is magnetically disordered at high temperature. Again, we take advantage of the crossings between $R_g^2/N$ curves at different values of $2N$ to pinpoint the phase transition between a compact and a swollen phase (see Fig.~\ref{fig:H05}c). We extrapolate the transition temperature $T^*$ in the $N\to\infty$ limit, obtaining $T^* = 2.56 \pm 0.02$. As before, we note that these findings align well with the MF prediction of a phase transition from a mixed-compact phase to a swollen one.

\subsubsection{$J_{11} = J_{22} =1$, $J_{12} = 2$, $H = 2.5$}

The MC results, shown in Fig.~\ref{fig:H25} confirm the presence of a compact-swollen phase transition. This is apparent in the behaviour of the intra-filament number of contacts per monomer, $c_{11} + c_{22}$, that decreases with increasing $T$ (see Fig.~\ref{fig:H25}a). Further, note that $c_{12} = 0$ for all values of $T$, highlighting the absence of a mixed phase at low $T$, as showcased by the compact segregated conformation in the snapshot at $T=1.9$. The presence of a transition is corroborated by the $T$-dependence of the variance of the contacts $Var(c) = Var(c_{11}+c_{22})$: these curves display a maximum that increases in height and moves towards higher temperatures as $2N$ increases (Fig. \ref{fig:H25}b). Further, as in previous cases, the presence of crossings in the $R_g^2/(2N)$ curves at different values of $N$ is used to pinpoint the location of the transition (see Fig.~\ref{fig:H25}c). We extrapolate the transition temperature in the thermodynamic limit, as previously detailed, to obtain $T^* = 3.15 \pm 0.02$.
Again, these findings align quantitatively with the MF predictions; it is worth mentioning that the estimated transition temperature is in good agreement with its mean-field counterpart $T^*_{MF} \approx 3.1$.

\subsubsection{$J_{11} = 1,J_{22} =0.5 $, $J_{12} = 2$, $H = 2.5$}

The MC results, shown in Fig.~\ref{fig:TADH25}, suggests the presence of a tadpole phase; this is highlighted by the difference between the values of $c_{11}$ and $c_{22}$, the number of intra-filament contacts per monomer, as well as their variance $Var(c_{11})$ and $Var(c_{22})$ (see Fig.~\ref{fig:TADH25}a,b). 
Indeed, both $c_{11}$ and $c_{22}$ decrease with increasing $T$, but the latter decreases faster. Moreover, the peaks in $Var(c_{22})$ consistently occur at lower temperature values. This behaviour indicates that the second filament, characterized by a weaker interaction strength ($J_{22}<J_{11}$), undergoes a transition to a swollen phase at lower temperatures than the first filament. This differential behavior in the transition temperatures facilitates the emergence of a 'tadpole' phase (see snapshots in Fig.~\ref{fig:TADH25}). A similar trend is observed by the crossings of the $R_{g,1}^2/{N}$ and $R_{g,2}^2/{N}$ curves  (see Fig.~\ref{fig:TADH25}c). We estimate the transition temperatures, that is, the boundaries of the tadpole region at fixed $H = 2.5$, by extrapolating the values of these crossings in the $N\to\infty$ limit. This gives,  $T_1^* = 3.0 \pm 0.2$ and $T_2^* = 2.1 \pm 0.3$. 
Also in this case, these findings align quite well with the predictions of the mean-field theory.

\section{Discussion and Conclusions}
\label{sec:Discuss}
In this work, we have formulated and analysed, by mean field theory and Monte-Carlo simulations, a minimal model of a magnetic diblock copolymer with ferromagnetic intra-filament and antiferromagnetic inter-filament couplings, and determined its equilibrium phase diagram. One specific aspect which we focussed on was understanding how an external magnetic field affects both the spatial organization and magnetic ordering of the system. The competition between like–like alignment within each block and antagonistic interactions across blocks introduces magnetic frustration that couples directly to polymer conformational degrees of freedom.

Within a Bragg–Williams mean-field framework, we derived the free-energy density as a function of five order parameters: the magnetizations of the two blocks, $m_1$ and $m_2$; their average densities, $\rho_1$ and $\rho_2$; and a mixing parameter, $\rho_{\rm mix}$, which quantifies the extent of spatial interpenetration. The mean-field theory predicts four distinct equilibrium phases are possible. First, there is a a swollen phase, where both filaments are extended and magnetically disordered. Second, we find a mixed compact phase, characterised by a single globule with oppositely magnetized, intertwined blocks. Third, a segregated compact phase is possible, consisting of two spatially separated globules that align with the external field. Finally, for asymmetric intra-block couplings, i.e. $J_{11}>J_{22}$, there exists a hybrid segregated (``tadpole'') phase, in which one block collapses into a magnetically ordered globule while the other remains swollen.

These mean-field predictions were tested against Monte Carlo simulations of magnetic diblock copolymers modeled as self-avoiding walks on a cubic lattice. 
We examined numerically different ``cuts'' of the phase diagram in the $H-T$ plane, either at fixed values of $H$ or at fixed values of $T$, and estimated the location of the phase transitions under those conditions. We further performed simulations both in the case $J_{11}=J_{22}$, that is, a symmetric diblock, or in the case $J_{11} \neq J_{22}$, confirming the presence of a tadpole phase. The Monte Carlo results show good agreement with the mean-field predictions, indicating that the Bragg–Williams approach captures the essential physics of frustration-driven conformational transitions in this heterogeneous magnetic polymer system.

\begin{figure}
    \centering
    \includegraphics[width=\linewidth]{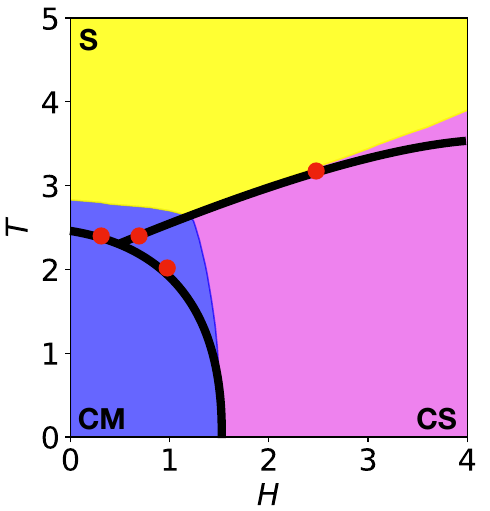}
    \caption{Sketch of the phase boundaries of the model of magnetic diblock copolymer on a cubic lattice (tick solid curves). These have been drawn according to four asymptotic ($N\to\infty$) MC estimates reported as red circles and discussed in the text. For comparison, these boundaries have been superimposed on the MF phase diagram (colored regions and corresponding boundaries) previously shown in Fig.~\ref{fig:Phases}}
    \label{fig:MC_phase_diagram}
\end{figure}

Overall, the system presented and studied in this work represents a minimal statistical-mechanical framework for understanding how competing internal interactions and external fields can control the morphology of magnetically heterogeneous polymers. It also constitutes a simple frustrated two-state polymer model in which antagonistic interactions couple directly to large-scale conformational organization. In the context of the biophysics of chromatin, we note that our model is also formally analogous to coarse-grained models of chromatin carrying antagonistic epigenetic marks -- such as those introduced in~\cite{michieletto2016polymer,owen2023,michieletto2019nonequilibrium,coli2019magnetic,jost2018,fiorillo2020modern,esposito2022polymer} -- where like states attract and unlike states interact unfavorably (either repulsively or sterically), leading to compartmentalization. In this analogy, the external magnetic field plays a role similar to a bias favoring one epigenetic state over the other -- for instance mimicking conditions in which active or inactive chromatin marks are globally promoted -- thereby shifting the balance between mixed and segregated conformations. Note that this biophysical link should be seen as qualitative, as our treatment is purely an equilibrium one and does not include epigenetic spreading dynamics. 

We hope that extensions of the model described in this work may serve as a useful starting point for investigating more complex responsive soft materials as well as antagonistic-state polymer systems in biological contexts.

\providecommand{\noopsort}[1]{}\providecommand{\singleletter}[1]{#1}%

\clearpage

\twocolumngrid 
\appendix
\onecolumngrid 

\begin{center}
\textbf{\large \vspace*{1.5mm} Field-induced phase transitions in ferro-antiferromagnetic diblock copolymers -- Supplemental Information } \\
\end{center}
\vspace*{5mm}

\section{Calculations for the mean-field approximation}
\label{sec:Fderivation}
Here we report in detail the calculations of the mean-field energy $E_{BW}^\gamma$ and the mean-field entropy $S_{BW}$ using the Bragg-Williams (or BW) approach.\\
Let $p(\{S\})$ be the probability of observing a certain spin configuration $S=\{S_{1}, S_{2}, \dots, S_{2N}\}$. Under the BW approximation, such a probability is written as the product of single-particle (spins) probabilities:
\begin{equation}
    p(\{ S\}) = \prod_{i = 1}^{{2N}} p^{1}(S_{i})
    \label{decomp}
\end{equation}
The magnetic field $H$ is homogeneous and the coupling constants ($J_{11}$, $J_{22}$ and $J_{12}$) are spin independent. Thus, naming $m_1$ and $m_2$ the average magnetization of the two filaments, we can write:
\begin{equation}
    p^{1}(S_{i}) = \frac{1 + m_{i}}{2} \delta_{S_{i}, +1} +  \frac{1 - m_{i}}{2} \delta_{S_{i}, -1}
    \label{singleSpinProb}
\end{equation}
where $m_{i} = m_{1}$ if $i \in \gamma_{1}$ and  $m_{i} = m_{2}$ if $i \in \gamma_{2}$. We can write a mean-field partition function of the magnetic interactions of the polymer using the probability $p$. Starting from the partition function, we get:
\begin{align*}
	\mathcal Z &= \sum_{\gamma \in \text{SAW}} \sum_{\{S\}} e^{-\beta \mathcal H(\gamma, \{S\})} \\
	&= \sum_{\gamma \in \text{SAW}} \sum_{\{S\}} p(\{S\}) e^{-\beta \mathcal H(\gamma, \{S\}) - \log p(\{S\})} \\
	& \ge \sum_{\gamma \in \text{SAW}} e^{-\beta \left \langle \mathcal H(\gamma, \{S\}) \right \rangle_{p} - \left \langle \log p(\{S\}) \right \rangle_{p}} \\
	&=  \sum_{\gamma \in \text{SAW}} e^{-\beta E^{\gamma}_{BW} + S_{BW}} = \mathcal Z_{BW}
\end{align*}
where in the third line we used Jensen's inequality. 
In this approximation, the energy reads:
\begin{align*}
E^\gamma_{BW} = \langle \mathcal H (\gamma , \{S\}) \rangle_p &= \left \langle -\frac{J_{11}}{2} \sum_{ij \in \gamma_{1}}S_{i}\Delta_{ij}^{\gamma}S_{j} -\frac{J_{22}}{2} \sum_{ij \in \gamma_{2}}S_{i}\Delta_{i,j}^{\gamma}S_{j} + 
J_{12} \sum_{i \in \gamma_{1}} \sum_{j \in \gamma_{2}}S_{i}\Delta_{i,j}^{\gamma}S_{j} - H \sum_{i \in \gamma}S_{i}  \right \rangle_p \\
&= -\frac{J_{11}}{2} \sum_{ij \in \gamma_{1}} \langle S_{i} \rangle \Delta_{ij}^{\gamma} \langle S_{j}\rangle  -\frac{J_{22}}{2} \sum_{ij \in \gamma_{2}} \langle S_{i} \rangle \Delta_{i,j}^{\gamma} \langle S_{j}\rangle  + 
J_{12} \sum_{i \in \gamma_{1}} \sum_{j \in \gamma_{2}}\langle S_{i}\rangle \Delta_{i,j}^{\gamma} \langle S_{j} \rangle - H \sum_{i \in \gamma} \langle S_{i} \rangle  \\
& = -\frac{J_{11} m_1^2}{2} \sum_{ij \in \gamma_{1}}\Delta_{i,j}^{\gamma}  -\frac{J_{22} m_2^2}{2}\sum_{ij \in \gamma_{2}}\Delta_{ij}^{\gamma} + J_{12} m_1 m_2 \sum_{i \in \gamma_{1} j \in \gamma_{2}} \Delta_{ij}^{\gamma}  - N H (m_1 + m_2) 
\end{align*}
where in the second equality we exploit the single spin decomposition of $p$ presented in Eq.~\eqref{decomp} and omit the subscript $\langle S_i\rangle_{p^1}$ on the averages of the spins. Notice that summations over subparts of the adjacency matrix of walk $\gamma$ appear in the expression of the mean-field energy $E^{\gamma}$. These terms are not trivial to compute because they depend on a particular SAW $\gamma$. To overcome this problem and simplify calculations, we rely on the so-called Hamiltonian walk approximation~\cite{garel1999phase, coli2019magnetic}, where we restrict the set of SAWs to the one that is space filling, that is, the Hamiltonian walks. A Hamiltonian walk is a walk that visits each vertex of a lattice embedded in a volume $V$ exactly once and has been used to study the equilibrium properties of highly compact polymers. For this kind of walk, the adjacency matrix of the SAW $\Delta$ takes the same form as the adjacency matrix of the underlying lattice and is characterized by the coordination number $z$. Here we consider $2N$-step configurations that, similarly to Hamiltonian walks, are contained in a volume $V$ but may in principle display a lower mean number of nearest neighbors $\rho z$, where $0 \le \rho \le 1$ is the density of the diblock copolymer. In this way the summation of all the entries of the adjacency matrix is:
\begin{equation}
\sum_{i,j}\Delta_{i,j} = 2 N z \rho
\label{HamApp}
\end{equation}
With the Hamiltonian walk approximation, the introduction of $\rho_{mix}$ and the splitting detailed in the main text, we obtain Eq.~\eqref{eq:Ebw} of the main text.\\
To compute the mean-field entropy $S_{BW}$ it can be useful to prove that the probability $p$ follows the identity $\langle \log p \rangle_p = 2N \langle  \log p_i \rangle_{p_i}$. 
In fact:
\begin{align*}
    S_{BW} =\langle \log p \rangle_p &= \sum_{\{S \}} p(\{S\})\log p(\{S\}) \\
    &= \sum_{\{S \}} \left ( \prod_{i = 1}^{2N} p^1(S_i) \right ) \sum_{i = 1}^{2N} \log \left (    p^1(S_i)\right ) \\
    &= \sum_{\{S \}} p^1(S_1)\dots p^1(S_{2N}) \log (p^1(S_1)) + p^1(S_1)\dots p^1(S_{2N}) \log (p^1(S_2)) + p^1(S_1)\dots p^1(S_{2N}) \log (p^1(S_{2N})) \\
    &= \sum_{i = 1}^{2N} \sum_{S_i = \pm 1} p^1(S_i) \log p^1(S_i) =  \sum_{i = 1}^{2N} 
    \langle \log  p^1 \rangle_{p^1} = 2N \langle \log  p^1 \rangle_{p^1}
\end{align*}
where in the third line we used $\sum_{S_i = \pm 1}p^1(S_i) = 1$. Using the expression of $p^1$ in Eq.~\eqref{singleSpinProb}, we immediately find Eq.~\eqref{entropyMF}.
\subsection{Mean-field Self-consistent Equations}
The self-consistent equations are derived from Eq.~\eqref{freeE} by minimizing the mean-field free energy density with respect to the five order parameters $m_{1}$, $m_{2}$, $\rho_{1}$, $\rho_{2}$, and $\rho_{mix}$:
\begin{subequations}
\begin{equation}
\frac{\partial f}{\partial m_1} = - \frac{H}{2}
- \frac{1}{2} J_{11} m_1 \rho_1 \left(1 - \frac{\rho_{mix}}{2} \right) z
+ \frac{1}{8} J_{12} m_2 \rho_{mix} z (\rho_1 + \rho_2)
+ \frac{1}{2} T \tanh^{-1}(m_1) = 0
\end{equation}
\begin{equation}
\frac{\partial f}{\partial m_2} =- \frac{H}{2}
- \frac{1}{2} J_{22} m_2 \rho_2 \left(1 - \frac{\rho_{mix}}{2} \right) z
+ \frac{1}{8} J_{12} m_2 \rho_{mix} z (\rho_1 + \rho_2)
+ \frac{1}{2} T \tanh^{-1}(m_2) = 0
\end{equation}
\begin{align}
\frac{\partial f}{\partial \rho_1} =&- \frac{1}{4} J_{11} m_1^2 \left(1 - \frac{\rho_{mix}}{2} \right) z
+ \frac{1}{8} J_{12} m_1 m_2 \rho_{mix} z
- \frac{\rho_{mix}}{(\rho_1 + \rho_2)\beta} \cr
&+ (1 - \rho_{mix}) \left[
    -\frac{1}{2 \rho_1 \beta}
    - \frac{(1 - \rho_1) \log(1 - \rho_1)}{2 \rho_1^2 \beta}
    - \frac{\log(1 - \rho_1)}{2 \rho_1 \beta}
\right] \cr
&- \frac{(2 - \rho_1 - \rho_2)\rho_{mix} \log\left(\frac{1}{2}(2 - \rho_1 - \rho_2)\right)}{(\rho_1 + \rho_2)^2 \beta}
- \frac{\rho_{mix} \log\left(\frac{1}{2}(2 - \rho_1 - \rho_2)\right)}{(\rho_1 + \rho_2) \beta} = 0
\end{align}
\begin{align}
    \frac{\partial f}{\partial \rho_2} = &- \frac{1}{4} J_{22} m_2^2 \left(1 - \frac{\rho_{mix}}{2} \right) z
+ \frac{1}{8} J_{12} m_1 m_2 \rho_{mix} z
- \frac{\rho_{mix}}{(\rho_1 + \rho_2)\beta} \cr
&+ (1 - \rho_{mix}) \left[
    -\frac{1}{2 \rho_2 \beta}
    - \frac{(1 - \rho_2) \log(1 - \rho_2)}{2 \rho_2^2 \beta}
    - \frac{\log(1 - \rho_2)}{2 \rho_2 \beta}
\right] \cr
&- \frac{(2 - \rho_1 - \rho_2)\rho_{mix} \log\left(\frac{1}{2}(2 - \rho_1 - \rho_2)\right)}{(\rho_1 + \rho_2)^2 \beta}
- \frac{\rho_{mix} \log\left(\frac{1}{2}(2 - \rho_1 - \rho_2)\right)}{(\rho_1 + \rho_2) \beta} = 0
\end{align}
\begin{align}
\frac{\partial f}{\partial \rho_{mix}} = & \frac{1}{8} J_{11} m_1^2 \rho_1 z
+ \frac{1}{8} J_{12} m_1 m_2 \rho_1 z
+ \frac{1}{8} J_{12} m_1 m_2 \rho_2 z
+ \frac{1}{8} J_{22} m_2^2 \rho_2 z \cr
&- \frac{(1 - \rho_1)\log(1 - \rho_1)}{2 \rho_1 \beta}
- \frac{(1 - \rho_2)\log(1 - \rho_2)}{2 \rho_2 \beta}
+ \frac{(2 - \rho_1 - \rho_2)\log\left(\frac{1}{2}(2 - \rho_1 - \rho_2)\right)}{(\rho_1 + \rho_2) \beta} = 0
\label{MFRhomix}
\end{align}
\end{subequations}
The stationarity condition with respect to $\rho_{mix}$ does not explicitly contain $\rho_{mix}$, implying that the free energy is affine in $\rho_{mix}$ when the other variables are fixed. As a result, the minimization of $f$ over $\rho_{mix} \in [0,1]$ generally produces a solution at the boundaries ($\rho_{mix} = 0$ or $\rho_{mix} = 1$) with the sign of $\partial f/\partial \rho_{mix}$ determining which boundary is selected.

\subsection{Mean-field approximation and Monte-Carlo simulations in the $H \to \infty$ limit}
\begin{figure}
    \centering
    \includegraphics[width=0.9\linewidth]{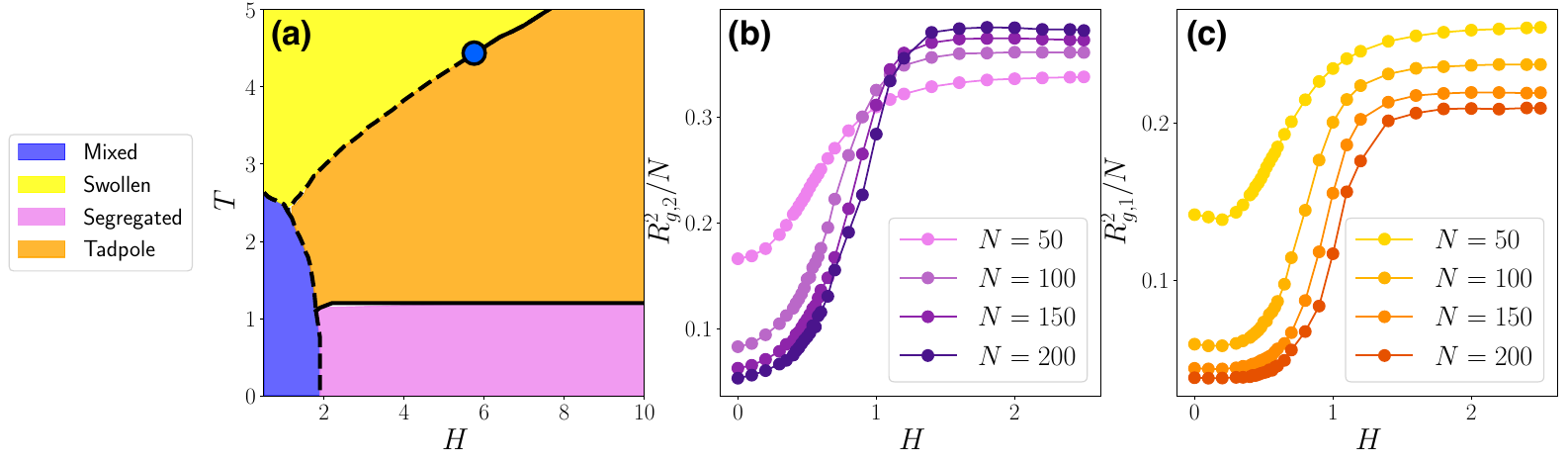}
    \caption{Mean-field phase diagram for $J_{11} = 1$, $J_{22} = 0.2$, $J_{12} = 2$ and Monte Carlo simulations for $T = 2.0$. In panel (a) we show the mean field phase diagram computed for $J_{11} = 1$, $J_{22} = 0.2$, $J_{12} = 2$. Notice how the boundary between the segregated and the tadpole phase becomes flat for large values of $H$. As in Fig. ~\ref{fig:Phases}, dashed and continuous lines denote respectively first and second order phase transitions whereas the blue circle denotes the tricritical point. In panel (b)-(c) we show the $R_{g,2}^2/N$ and $R_{g,1}^2/N$ trends for $N = 50,100,150,200$. In the case of filament 2 (panel (b)), for $H \gtrsim 1.2$ the curves corresponding to large values of $N$ lie above the others, as expected from a swollen polymer. The opposite behavior is observed in panel (c) since filament 1 is instead in a collapsed phase. }
    \label{fig:tad02}
\end{figure}

When $H \to \infty$ one expects that all the monomers carry $+1$ spin variables. In this limit, we expect to recover the so called ISAW (interacting self avoiding walk), a model for the $\theta$ point transition where all the monomers of the polymer attract each other whenever they are nearest neighbors in space. Given a lattice with coordination number $z$ and a coupling constant $J$, according to this model the $\theta$ point is located at a temperature $T_\theta = Jz$. 

We know that for a large magnetic field, the two block of the magnetic diblock copolymer behaves as independent magnetic polymers. For the block $b$, the Hamiltonian is:
\[
\mathcal H_b = -\frac{J_{bb}}{2}\sum_{i,j}S_i \Delta_{ij}S_j - H \sum_{i}S_i
\]
where $b = 1,2$. The free energy per monomer of filament $b$ is reported in Eq. 11 of Ref.~\cite{garel1999phase}:
\[
f_b = -\frac{1}{\beta}\left (\log \frac{z}{e} + \log (2 \cosh (\phi_b + \beta H)) - \frac{1-\rho_b}{\rho_b}\log(1 - \rho_b) \right ) + \frac{\phi^2_b}{2 \beta^2\rho_b J_{bb} z}
\]
where $\rho_b$ is the density of block $b$ and $\phi_b = \beta J_b z \rho_b m_b$ is a field proportional to the magnetization per monomer $m_b$. Performing a virial expansion of the osmotic pressure $\Pi(\rho_b) = \rho_b^2 \partial f_b/\partial \rho_b$ one can write down the second virial coefficient:
\[
B_{2, b}(\beta, H) = \frac{1}{2\beta} - \frac{1}{2}J_{bb}z \tanh^2(\beta H)
\]
that, for $H \to \infty$, simplifies to:
\[
B_{2, b}(\beta, H\to \infty) = \frac{1}{2\beta} - \frac{1}{2}J_{bb}z 
\]
which vanishes when $T^*_b = T_{\theta,b} = J_{bb} z$ (the third virial coefficient is positive at $T = T^*$). As a consequence, the transition line between the compact segregated and the swollen segregated phases in the symmetric case $J_{11} = J_{22}$ (and the lines between compact segregated, tadpole and swollen segregated phases in the asymmetric case $J_{11} \neq J_{22}$) become flat in the $H \to \infty$ limit, converging to $T_{\theta, b} = J_{bb}z$.

Let us consider the asymmetric case $J_{11} = 1$, $J_{22} = 0.2$ and $J_{12} = 2$. The two mean-field critical temperatures in the $H \to \infty$ limit are $T_{\theta, 1} = 6.0$ and $T_{\theta, 2} = 1.2$. In Fig. \ref{fig:tad02}(a) we show the mean-field phase diagram corresponding to these parameter choice. Notice that, as expected, the boundary between the segregated and the tadpole phase is flat for great magnitudes of the magnetic field. Monte Carlo simulations performed at $T = 2.0$ are coherent with these picture. In Fig. \ref{fig:tad02}(b) and  \ref{fig:tad02}(c) we show respectively the trends of $R_{g,2}^2/N$ and $R_{g,1}^2/N$. For $H \gtrsim 1.2$ both trends become flat, as expected from the mean-field prediction. In the first case we expect the filament $2$ to be swollen, and in fact the high $N$ curves lie above the low $N$ ones; on the other hand, the filament $1$ is expected to be collapsed, leading the low $N$ curves to lie below the high $N$ ones.

\section{Diagnostics of the Monte Carlo simulations}
\begin{figure}
    \centering
    \includegraphics[width=0.8\linewidth]{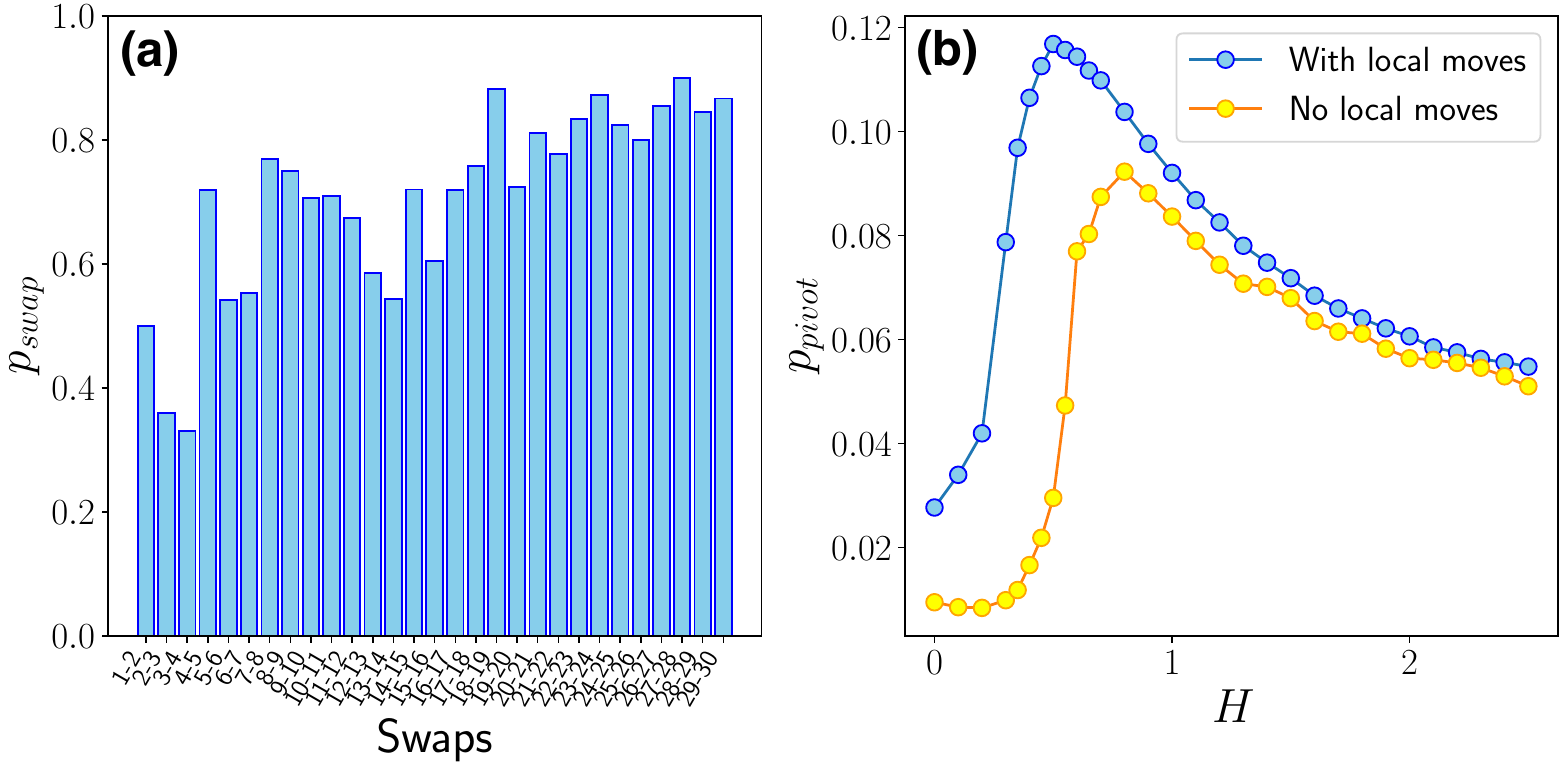}
    \caption{{Swap and pivot probability for simulations at $T = 2.5$ ($J_{11} = J_{22} = 1$ and $J_{12} = 2$).} Panel (a): histogram of the swap probability, referring to the simulations reported in Sec.~\ref{FixedT125}. The labels on the $x$-axis denote the indices of the adjacent Markov chains between which a swap is attempted. The magnetic field values corresponding to these labels are listed in Tab.~\ref{tab1}. No bottlenecks appear in the histogram, thus configurations are free to move across the whole set of magnetic fields. Panel (b): acceptance probability of pivot moves when either local moves are present (blue dots) or absent (yellow dots). At low values of $H$, when the system is in the mixed compact phase, the acceptance probability of pivot moves is lower in athe bsence of local moves ($p_{pivot} \simeq 7 \cdot 10^{-3}$).  }
    \label{fig:diagnostics}
\end{figure}

\label{AppendixDiagnostics}
In this appendix, we report some technical details regarding the Monte Carlo simulations discussed in the main text. In particular, we focus on the simulations performed at fixed temperature $T = 2.50$, with $J_{11} = J_{22} = 1$ and $J_{12} = 2$ and different values of the magnetic field $H$.  

As written in section \ref{sec:MonteCarlo}, to speed up the convergence to equilibrium and reduce the autocorrelation times, we implemented the Multiple Markov Chains algorithm, running several Markov chains in parallel and allowing chains that run at adjacent values of $H$ or $T$ to try swapping their configurations with a certain frequency. In a good MMC sampling, the swap probability should be comparable on all possible chains to allow the configurations to explore all possible chains. Moreover, in ~\cite{10.1063/1.1831273}, the authors show that an acceptance ratio of $0.20$ or greater yields the best possible sampling performance. Importantly, no bottlenecks, i.e., one or more Markov chain pairs with a swap probability significantly smaller than the others, should appear in the system to avoid an ergodicity breaking in the global Markov chain made of all the single MC chains. 
In our simulations, we set a trial swap frequency of $10^4$ MC steps. Data have been collected in every $10^3$ MC step.
In Fig. \ref{fig:diagnostics}(a) we show the histogram of the swap probabilities: all of them are greater that $0.30$. For strong magnetic fields ($H > 1.2$) the swap probabilities become quite large. This is because the distribution of the magnetization for such Markov Chains are highly superimposed. However, no critical bottlenecks are observed in the whole range of magnetic fields; the same is true when using swapping Markov chains at different values of $T$. In Tab.~\ref{tab1} we show the Pearson correlation coefficient between adjacent Markov Chains. Given two time series of an observable $X$, computed at $H_i$ and $H_{i+1}$ and made of $n$ measures each, the Pearson coefficient between these two observables is given by:
\[
\rho_{X}(i, i+1) = \frac{\sum_{a = 1}^n (x_a^i - \bar x^i)(x_a^{i+1} - \bar x^{i+1})}{\sqrt {\sum_{i = 1}^n (x_a^i - \bar x^i)^2  \sum_{i = 1}^n (x_a^{i+1} - \bar x^{i+1})^2 }}
\]
where $a$ is an index that runs all measures, $x_{a}^{i}$ is the $a$-th measure of $x$ in the $i$-th chain, and $\bar x^i = 1/n \sum_{a = 1}^n x_a^i$. When $|\rho| \simeq 1$, two variables are perfectly linearly correlated. On the other hand, if $|\rho| \simeq 0$ then two variables are not linearly correlated. The largest Pearson coefficient observed in Tab. \ref{tab1} is $\rho_E(1,2)=0.087$, calculated between the energy time series obtained at $H_1 = 0.00$ and $H_2 = 0.10$. We can therefore assume that our trial swap frequency is sufficiently high to avoid (linear) correlations between adjacent chains.

In Tab. \ref{tab1} there are also the integrated autocorrelation times for the energy, magnetization, contacts, and gyration radius, computed for every Markov chain (the unit of time is $10^3$ MC steps). 
We further compare these results with the case where no local moves are performed (the corresponding autocorrelation times are reported in Tab.~\ref{tab2}). 
For all observables and almost all the $H$ values, the autocorrelation times in Tab.~\ref{tab2} are orders of magnitude larger than the ones reported in Tab.~\ref{tab1}. Indeed, the pivot acceptance probability is reduced without local moves: see, for instance, Fig. \ref{fig:diagnostics}(b) where $p_{pivot}$ is one order of magnitude lower without local moves with respect to the case where local moves are present. 

\begin{table}[h]
    \centering
    \begin{tabular}{| c || c | c | c | c || c | c | c | c |}
        \hline
        $H$ & $\tau_E$ & $\tau_M$ & $\tau_C$ & $\tau_{RG}$ & $\rho(E)$ & $\rho(M) $ & $\rho(C)$ & $\rho(R_G)$\\
        \hline
        $0.00$ & $120(50)$ & $2.7(3)$ & $120(50)$ & $130(50)$ & $0.087$  & $0.010$ & $0.010$ & $0.019$ \\
        $0.10$ & $160(70)$ & $29(9)$ & $140(40)$ & $110(30)$ & $0.079$  & $0.010$ & $0.010$ & $0.019$ \\
        $0.20$ & $190(70)$ & $54(20)$ & $190(70)$ & $180(70)$ & $0.050$ & $0.013$ & $0.018$ & $0.037$ \\
        $0.30$ & $210(90)$ & $60(20)$ & $200(80)$ & $100(30)$ & $0.062$ & $0.005$ & $0.002$ & $0.005$ \\
        $0.35$ & $140(40)$ & $50(10)$ & $150(50)$ & $70(20)$ & $0.013$ & $0.021$  & $0.014$ & $0.020$ \\
        $0.40$ & $80(20)$ & $29(6)$ & $90(20)$ & $31(6)$ & $0.015$ & $0.012$ & $0.019$ & $0.067$ \\
        $0.45$ & $60(10)$ & $23(5)$ & $70(20)$ & $23(4)$ & $0.012$ & $0.008$ & $0.001$ & $0.045$ \\
        $0.50$ & $13(2)$ & $3.4(3)$ & $17(3)$ & $4.1(4)$ & $0.006$ & $0.004$ & $0.047$ & $0.001$ \\
        $0.55$ & $11(1)$ & $2.9(2)$ & $15(2)$ & $4.4(4)$ & $0.005$ & $0.016$ & $0.039$ & $0.011$ \\
        $0.60$ & $3.2(3)$ & $1.32(6)$ & $4.7(4)$ & $1.9(1)$ & $0.006$ & $0.034$ & $0.040$ & $0.005$ \\
        $0.65$ & $2.4(2)$ & $1.46(9)$ & $3.4(2)$ & $1.7(1)$ & $0.005$ & $0.016$ & $0.013$ & $0.014$ \\
        $0.70$ & $1.46(6)$ & $1.23(5)$ & $2.2(1)$ & $1.35(6)$ & $0.009$ & $0.007$ & $0.016$ & $0.002$ \\
        $0.80$ & $1.43(6)$ & $1.22(6)$ & $1.80(9)$ & $1.25(5)$ & $0.014$ & $0.001$ & $0.006$ & $0.001$ \\
        $0.90$ & $2.21(7)$ & $1.34(8)$ & $3.55(4)$ & $1.28(5)$ & $0.005$ & $0.011$ & $0.007$ & $0.001$ \\
        $1.00$ & $1.46(1)$ & $1.08(4)$ & $2.05(1)$ & $1.43(8)$ & $0.009$ & $0.009$ & $0.001$ & $0.001$ \\
        $1.10$ & $1.50(2)$ & $1.09(4)$ & $2.10(1)$ & $1.48(8)$ & $0.001$ & $0.009$ & $0.001$ & $0.001$ \\
        $1.20$ & $1.68(1)$ & $1.11(4)$ & $2.18(1)$ & $1.39(7)$ & $0.005$ & $0.005$ & $0.009$ & $0.017$ \\
        $1.30$ & $1.84(1)$ & $1.15(4)$ & $2.61(1)$ & $1.53(8)$ & $0.004$ & $0.005$ & $0.009$ & $0.017$ \\
        $1.40$ & $1.83(3)$ & $1.16(5)$ & $2.65(1)$ & $1.51(7)$ & $0.014$ & $0.006$ & $0.008$ & $0.006$ \\
        $1.50$ & $1.72(2)$ & $1.11(4)$ & $2.38(1)$ & $1.58(7)$ & $0.018$ & $0.005$ & $0.009$ & $0.017$ \\
        $1.60$ & $1.91(7)$ & $1.09(3)$ & $2.74(2)$ & $1.81(7)$ & $0.012$ & $0.008$ & $0.019$ & $0.032$ \\
        $1.70$ & $2.37(1)$ & $1.12(4)$ & $3.49(1)$ & $1.87(8)$ & $0.010$ & $0.006$ & $0.009$ & $0.011$ \\
        $1.80$ & $2.04(4)$ & $1.12(4)$ & $2.79(2)$ & $1.88(1)$ & $0.011$ & $0.010$ & $0.002$ & $0.001$ \\
        $1.90$ & $2.19(2)$ & $1.09(3)$ & $3.15(2)$ & $1.77(1)$ & $0.025$ & $0.020$ & $0.012$ & $0.017$ \\
        $2.00$ & $2.08(2)$ & $1.12(3)$ & $3.02(1)$ & $1.77(1)$ & $0.010$ & $0.006$ & $0.009$ & $0.040$ \\
        $2.10$ & $2.20(1)$ & $1.03(4)$ & $3.73(2)$ & $1.99(6)$ & $0.006$ & $0.020$ & $0.003$ & $0.032$ \\
        $2.20$ & $2.16(3)$ & $1.07(3)$ & $3.05(3)$ & $1.80(6)$ & $0.001$ & $0.014$ & $0.009$ & $0.013$ \\
        $2.30$ & $2.32(1)$ & $1.07(4)$ & $3.10(1)$ & $1.97(2)$ & $0.006$ & $0.012$ & $0.015$ & $0.020$ \\
        $2.40$ & $2.68(1)$ & $1.11(4)$ & $3.83(1)$ & $2.16(4)$ & $0.006$ & $0.003$ & $0.012$ & $0.017$ \\
        $2.50$ & $2.34(9)$ & $1.05(3)$ & $3.18(3)$ & $1.95(4)$ & $-$ & $-$ & $-$ & $-$ \\
        \hline
    \end{tabular}
\caption{Integrated autocorrelation times and Pearson coefficients for the simulations described in Section \ref{T250MC}. The first four columns report the autocorrelation times of the energy, magnetization, contacts, and radius of gyration for the MC simulations performed at $T = 2.50$. Notice that we collected data every $10^3$ MC sweeps, thus this number is the time unit. Errors on the integrated autocorrelation times are computed as $\sigma_{\tau} = \tau \sqrt{(4W + 2)/N}$ where $N$ is the length of the time series and $W$ is the cutoff imposed to compute $\tau$. The second four columns contain the Pearson correlation coefficient for the same observables computed on two consecutive values of the magnetic field (so that the values on the first row are related to the correlation between $H = 0.00$ and $H = 0.10$).}
    \label{tab1}
\end{table}

\begin{table}[h!]
\centering
\begin{tabular}{| c || c | c | c | c |}
\hline
$H$ & $\tau_E$ & $\tau_M$ & $\tau_C$ & $\tau_{R_g}$ \\
\hline
0.00 & 500(300) & 150(70) & 2100(1700) & 700(400) \\
0.10 & 300(100) & 20(50) & 1300(900) & 400(200) \\
0.20 & 200(80) & 140(40) & 500(200) & 300(100) \\
0.30 & 900(700) & 300(100) & 1400(1000) & 500(200) \\
0.35 & 500(200) & 400(100) & 700(300) & 500(200) \\
0.40 & 3000(3000) & 2000(2000) & 3000(3000) & 2000(2000) \\
0.45 & 3000(3000) & 3000(3000) & 3000(3000) & 2000(1000) \\
0.50 & 4000(4000) & 4000(5000) & 4000(4000) & 3000(2000) \\
0.55 & 6000(6000) & 5000(5000) & 7000(7000) & 5000(6000) \\
0.60 & 5000(6000) & 4000(4000) & 6000(6000) & 4000(5000) \\
0.65 & 5000(5000) & 3000(4000) & 6000(6000) & 4000(4000) \\
0.70 & 4000(4000) & 3000(3000) & 5000(4000) & 3000(3000) \\
0.80 & 1500(900) & 1000(700) & 2000(1000) & 900(600) \\
0.90 & 3000(3000) & 1000(500) & 3000(4000) & 2000(2000) \\
1.00 & 1200(900) & 600(300) & 2000(1000) & 700(600) \\
1.10 & 1600(1500) & 700(300) & 2000(2000) & 900(700) \\
1.20 & 2000(2000) & 800(400) & 3000(3000) & 1100(1100) \\
1.30 & 800(400) & 500(200) & 2000(2000) & 400(200) \\
1.40 & 700(400) & 400(100) & 900(500) & 300(200) \\
1.50 & 1300(1300) & 200(40) & 1200(800) & 500(300) \\
1.60 & 1500(1400) & 300(100) & 2000(2000) & 500(400) \\
1.70 & 1100(900) & 140(30) & 1300(1000) & 300(200) \\
1.80 & 400(200) & 200(50) & 600(400) & 300(200) \\
1.90 & 800(600) & 200(40) & 1100(900) & 500(300) \\
2.00 & 300(200) & 70(10) & 500(300) & 100(60) \\
2.10 & 200(80) & 80(10) & 300(100) & 80(30) \\
2.20 & 300(100) & 60(10) & 300(100) & 90(30) \\
2.30 & 1100(900) & 40(90) & 1400(1000) & 300(300) \\
2.40 & 2000(2000) & 20(30) & 3000(3000) & 600(500) \\
2.50 & 2000(2000) & 5(60) & 3000(3000) & 800(800) \\
\hline
\end{tabular}
\caption{Integrated autocorrelation times of energy, magnetization, contacts, and radius of gyration for the simulations described in Section \ref{T250MC} \emph{without} local moves. The time unit is $10^3$ MC sweeps. For almost every value of $H$, the integrated autocorrelation times and their errors are at least one order of magnitude greater than the corresponding times obtained by performing local moves.}
\label{tab2}
\end{table}

\newpage
\section{Characterization of the phases with KMeans}
When in the compact-segregated phase, a diblock copolymer conformation is characterized by two globular filaments, with a negligible number of inter-filament contacts. Because of this fact, given a polymer configuration in this phase, it should be possible to classify the monomers from the different blocks into two clusters with an inference scheme, such as an unsupervised learning algorithm.
We perform this task exploiting a well-known unsupervised classification algorithm, the K-means method~\cite{rokach2005clustering}, that classifies points in a preselected number $n$ of clusters based on a certain metric. In our case, we seek $n = 2$ clusters and, since monomers live in a 3D Euclidean space, we classify clusters based on the ordinary Euclidean distances. 

\begin{figure}
    \centering
    \includegraphics[width=0.5\linewidth]{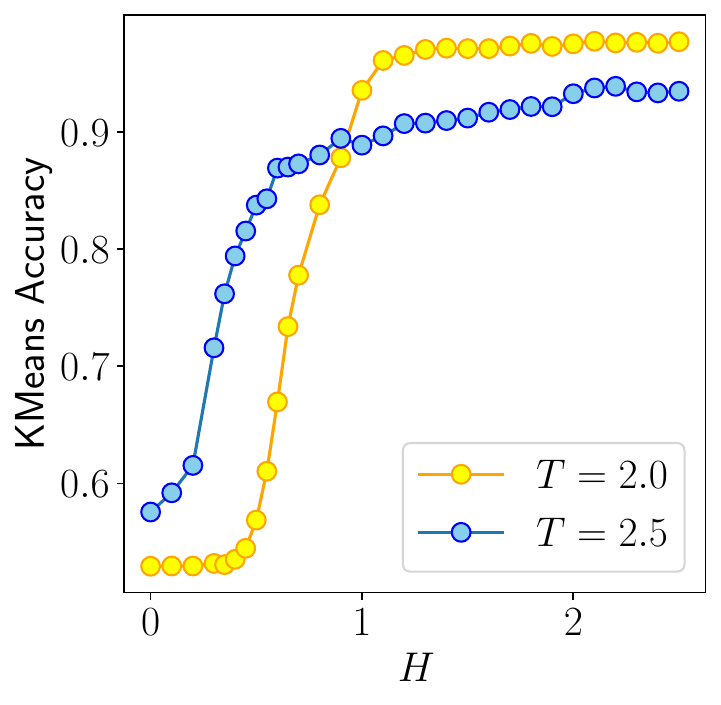}
    \caption{Accuracy of the KMeans algorithm applied to the classification of the monomers from the conformations produced via MC simulations at $T = 2.0$ and $T = 2.5$ ($J_{11} = J_{22} = 1$). The accuracy increases monotonically as a function of $H$: when $H \simeq 0$, the diblock polymer is in the mixed compact phase, and the classification is practically random, while,  when $H \gg 0$, compact segregated conformation as easy to label. 
    }
    \label{fig:KMeans}
\end{figure}

In Fig.~\ref{fig:KMeans} we show the results of such classifications for the configurations sampled during the simulations described in Sections~\ref{FixedT125} and \ref{T250MC}. We define the accuracy of the algorithm as the ratio between the number of correctly classified monomers and the total number of monomers; we report the results for $2N = 400$, with smaller sizes showing similar results. When $H \simeq 0$, the accuracy of the algorithm is $\approx 50\%-60\%$: this is expected since the diblock copolymer collapses in a mixed compact phase, i.e., a single cluster when the external magnetic field is weak. As such, it is impossible to correctly label the monomers, and the result is close to what would be obtained with a random assignment. In the opposite limit $H \gg 0$, the system is in a compact segregated phase, and the algorithm works very well, with an accuracy of $ \approx 96\%$ for $T = 2.0$ and $\approx 93\%$ for $T = 2.5$. This slight difference arises from a light swelling of the polymer conformations at $T = 2.5$, as compared to the $T=2.0$ case, caused by the slight increase in temperature.

\end{document}